\documentclass{article}

\usepackage{amssymb}

\usepackage{epsfig}

\usepackage{lscape}

\usepackage{graphicx,psfrag,amsmath}

\usepackage{yfonts}

\usepackage{xcolor}

\def\t{temperature }
\def\tn{temperature}

\def\zz{neutrino }
\def\zzn{neutrino}
\def\zzs{neutrinos }
\def\zzsn{neutrinos}

\def\be{\begin{equation}}
\def\ee{\end{equation}}
\def\bea{\begin{eqnarray}}
\def\eea{\end{eqnarray}}

\def\simlt{\lower.5ex\hbox{$\; \buildrel < \over \sim \;$}}
\def\simgt{\lower.5ex\hbox{$\; \buildrel > \over \sim \;$}}
\def\simpropto{\lower.2ex\hbox{$\; \buildrel \propto \over \sim
\;$}}

\newcommand{\eq}[1]{Eq\,\ref{#1}}

\newcommand{\tend}[1]{$10^{#1}$}
\newcommand{\tennd}[1]{10^{#1}}
\newcommand{\nrd}[2]{${#1}\times{10^{#2}}$}
\newcommand{\nrnd}[2]{({#1}\times{10^{#2}}\,)}

\newcommand{\tx}[1]{$t_{#1}$}
\newcommand{\txnd}[1]{t_{#1}}

\def\dtn{detection }

\def\pho{photon }
\def\phon{photon}
\def\phos{photons }
\def\phosn{photons}

\def\T{temperature }

\def\t{time }
\def\tn{time}

\def\sy{system } 
\def\sys{systems }
\def\syn{system}

\def\yy{energy }
\def\yyn{energy}
\def\yys{energies }
\def\yysn{energies}

\def\pl{particle }
\def\pls{particles }
\def\pln{particle}
\def\plsn{particles}

\def\sctg{scattering }
\def\sctgn{scattering}

\def\prob{probability }

\def\hf{\tfrac{1}{2}}

\def\zz{neutrino }
\def\zzn{neutrino}
\def\zzs{neutrinos }
\def\zzsn{neutrinos}

\def\gvtl{gravitational }

\def\bh{black hole }

\def\bhs{black holes }

\def\csss{cross-section }
\def\csssn{cross-section}

\def\beq#1\eeq{\begin{equation}#1\end{equation}}
\def\beql#1#2\eeql{\begin{equation}\label{#1}#2\end{equation}}

\def\bea#1\eea{\begin{eqnarray}#1\end{eqnarray}}
\def\beal#1#2\eeal{\begin{eqnarray}\label{#1}#2\end{eqnarray}}

\def\bu{burst }
\def\bun{burst}
\def\bus{bursts }
\def\busn{bursts}

\def\eb{{\bar E^\nu}}
\def\ecmb{E^\nu_{cmb}}
\def\rsh{redshift }
\def\rshn{redshift}

\def\em{emission }
\def\emn{emission}

\def\sp{spectrum }
\def\spn{spectrum}

\def\rsln{resolution}

\def\anph{annihilation photon }
\def\anphn{annihilation photon}

\begin{document}

\title{Signals of Bursts from the Very Early Universe }

\author{Leo Stodolsky\\
Max-Planck-Institut f\"ur Physik,\\ Boltzmannstr.\,8,
85748 Garching, Germany
\\
\\
Joseph Silk\\
Institut d'Astrophysique de Paris, UMR7095:CNRS \\
 UPMC-Sorbonne University, F-75014, Paris, France \\\\
 Dept. of Physics and Astronomy,
The Johns Hopkins University\\
3400 N. Charles Street
Baltimore, Maryland 21218, USA \\\\
 BIPAC, University of Oxford, 1 Keble Road,
Oxford, OX1 3RH, UK\\
} 

\maketitle
\begin{abstract}
We consider possible observable signals from explosive events in the very
early universe, ``\bus". These could   be  expected in connection
 with massive black hole  or
``baby universe'' formation. We anticipate that such
major disruptions of spacetime would be associated with 
 neutrino  and perhaps other  pulses. While these seem to be not
 detectable  directly, we discuss how they could
lead to potentially  observable signals.
We analyse how the pulses from  very early times
  may  ``escape'',  that is
  propagate to  the last scattering epoch at the time  $t_{cmb}$ and later,
    or alternatively be  absorped  earlier,
``contained''.  The possibly detectable signals include effects 
on small regions of  the CMB, a  soft
x-ray resulting from positron prodution, or a nonthermal addition to
 the  relic neutrino background.
\end{abstract}

\section{Introduction}
Some years ago \cite{jl}, we contemplated the possibility
 that in the very early
universe, explosive events could take place, analogous 
 to the supernovas
seen in the presently observable universe. These might be induced by such
processes as the collapse of massive  regions to \bhs
or  the formation of ``baby universes''. Although such events would  lead
to regions of spacetime which are   physically disconnected from us,
 it is plausible that during their formation, peripheral or transient
phenomena occur, as is familiar  for  the optical or  \zz bursts
 accompanying  supernovae. Just as for the supernovae,  a quiescent remnant
may be left behind, while  a dramatic explosive effect reaches the ``outer
world''.
Components of 
 such \bus could  reach us through their weakly interacting \pls \cite{flt}, 
 and  in \cite{jl} we
considered some of the basic  features in terms of 
 local observations on or near the earth.

 To set the context,  the possibility  of baby universes was predicted from calculations of the wave function of the universe
  \cite{Hartle:1983ai}, and   their actual formation was proposed 
 by \cite {Hawking:1988wm}. They are 
  theoretically motivated by string theory, and more 
 specifically by models such as eternal inflation
 \cite{Guth:2000ka},  and more recently by discussions
 such as \cite {Dijkgraaf:2005bp, Hebecker:2018ofv}.
 Observational signatures have been studied of bubble universe collisions, most notably via
 searching for hot spots in the CMB sky \cite{Feeney:2010jj}.
 Bubble universe creation is an ongoing phenomenon in 
 the large number  of scalar fields   in the generic
 inflationary landscape \cite{Easther:2016ire}.

  Another formation  channel for
 baby universes is associated with the density
 fluctuation amplitude threshold for PBH formation \cite{Musco:2020jjb}.
 Exceed this and one produces baby universes
 \cite{Carr:1974nx, Kopp:2010sh}.  This idea couples the rate and abundance of baby
 universe creation events to the PBH  formation history and formation rate. 
 This adds credibility to the baby universe hypothesis,
 since PBHs provide  a plausible scenario for
 dark matter with a number of potentially
 significant probes under development,
 most notably with next generation gravitational wave observatories, 
  
{ Observation or \dtn  of such events would evidently  open a new chapter
in observational cosmology. 
However, } there is
a great difficulty in directly detecting the existence of such energetic 
but very early time events, namely the  high redshift  to be
anticipated. The \pls in the \bun, such as \zzsn, will arrive at
the earth with an \yy $a (t_{em})E_{em}$, where $E_{em}$ is the \yy 
in the rest frame of their
emission, and $a(t_{em})$ is the cosmological
expansion parameter at the time of the burst.
Thus \zzs emitted from an  event at cosmic time $t_{em}= 1$ second
 will have their \yys reduced by a factor $\nrnd{2}{-10}$ at the earth. Since
the \zz \csss for \dtn  is strongly \yy dependent,
 this would  present formidable 
difficulties, even if we find  \cite{jl}  that the flux  factor stops
 decreasing for very early  \em  times.
 
Here we would like to consider a less direct but perhaps more
feasible approach towards  finding evidence of such
 ``early \bus'', namely
effects from the \bus at very early timss or high \rsh but neverthelss
leading to a possibly observable signal.

We shall consider three possible such signals, involving different
technologies.
One would be the presence of positrons, created by \bu \zzsn. These could
  give an observable soft x-ray signal \cite{xrs}.   

Another  is  a ``heat signal'' concerning the \yy injected
by absorped  \bus into small regions on the CMB. We note
 that there is a rich literature on early energy
 injection into the CMB. This was pioneered by \cite{Hu:1992dc}
and developed in many more recent studies, culminating in elegant theoretical
 formalisms  including \cite{Acharya:2023wax},
 in ground-breaking experiments  such as COBE FIRAS \cite{Fixsen:1997xq}
 and in proposals for future CMB SD experiments \cite{Chluba:2019nxa}.
 Here we will use a  simplified method, adapted to
a qualitative discussion of the hypothesized \busn  .

Finally, there is a question involving relic \zzsn.
An important distinction concerning the \bus turns out to be
if they can ``escape'', propagate to 
late \tn s, or be ``contained '',  absorped at early \tn s. 
The    ``escape'' to present \tn s of very low \yy
\zzs (or other weakly interacting particles) would lead to another,
 but very difficult to observe  signal: a 
nonthermal  component of relic \zzsn. 
  This would involve new technologies
 for the  observation of low \yy
neutrinos \cite{Rossi:2024cxa} and the determination of their \spn.

 This paper is at
the interface of \pl physics and astrophysics.
We have chosen a
 style, frequently involving  intuitive  and approximate
arguments, which we hope has enabled us to identify the important issues
of this new subject. We trust this has made 
 the  discussion easily accessible to  both communities.

\section{Neutrino Propagation}

The \zz is the only known \pl at present with purely weak interactions,
and so it can travel relatively ``far''. 
In a core-collapse supernovae almost all the \yy is carried away by \zzsn.
Although practically all types of \pls are involved in the event and even the \zzs
themselves are in a transitory thermal equilibrium (the ``neutrinosphere''),
their  weak interactions imply that they are the ``first to leave'' and
carry most of the \yyn. 
We imagine that something analogous  takes place in the  case of ``horizon collapse''
 and  that the \zzs carry most of the
outgoing  \yyn.
However the \yy \sp of the individual \zzs is a complicated matter, 
 and  we simply  examine some illustrative
 models, hoping to extract some general features.

Behind the \zz pulse, there can  be more strongly interacting components,
with  electromagnetic or strong interactions. This would give rise to a 
localized heating, whose effects can also contribute to  a ``heat
signal'' discussed below.

\subsection{Neutrino mean-free-path}\label{mfb}
The distance traveled by a \zz before it interacts is obtained from
 the mean-free-path $\lambda $, which is given by $\lambda =1/(\rho
\sigma)$, where $\rho$ is the density of  scatterers and $\sigma$ the
 \csss for the \zz on the scatterer. We will take this to be a generic 
weak interaction cross-section,  whose most salient feature is its increase with
\yyn. This leads to much greater containment for high \yy \pls vis-a-vis
low \yy ones (See Table 1).

 The mean-free-path refers to a physical spatial distance and so should
be used in conjunction with the proper distance. However in the standard
cosmological spacetime metric that
 we use, $ds^2=dt^2 - (a(t){\bf dx})^2$, one has
for essentially massless \pls like \zzs that $dt=a(t) dx$. Therefore when we
write for the  decay of a pulse $\sim - \lambda(t) dt$, we are in fact
using the correct proper distance. Note that
 since we use $c=1$ units (and generally natural units)  ,  $\lambda $ is also the time between interactions
for essentially massless
 \plsn.

 The \csss will of course vary somewhat with the scatterer, but
for our present semi-quantitative purposes, we take a generic weak
interaction \csss
\beql{cs}
\sigma\approx \biggl(\frac{\alpha}{M^2}\biggr)^2S\,,
\eeql
with $\alpha=1/137, M=100\, \rm GeV$ and $S$ the center-of-mass
 \yy squared of the \sctg .

{
\eq{cs} results from the exchange of W and Z bosons in the
 electroweak
standard model \cite{hollik}. The scatterers may be leptons, quarks or other 
\pls interacting with \zzs via such exchanges.
A salient feature of \eq{cs} is its \yy dependence, which as will be seen
below, leads to a strong \T and \yy dependence of various aspects of the
problem.

 If $E^\nu(t)$ is the \yy of the \zz and $E^i(t)$ 
the \yy of a scatterer, one has, averaged over directions,
$S\approx E^\nu(t)E^i(t)$. This follows from $S= (p^\nu-p ^i)^2$
where the $p$ are the 4-momenta of a \sctg partner $i$ and the \zzn.
. This leads to
the 4-product $S\approx 2 p^\nu\cdot p^i= 2E^\nu E^i(1-cos\theta)$, 
with $\theta$
the angle between the \zz and the \sctg partner $i$. After an angular
average the $cos\theta$ term drops out and  one has
 $S\approx E^\nu(t)E^i(t)$ where, as
throughout, we work only
to order-of-magnitude accuracy. }

 At  \T $T$ with the \yy for a scattering partner $\sim T$ one  will 
have  $S\sim E^\nu(t)T(t)$.
For the number density of \pls $\rho$ that one has   for relativistic radiation,
\beql{ro}
\rho \approx  T^3,
\eeql
one thus arrives at
\beql{lam}
\lambda \sim \frac{M^4}{E^\nu T^4\alpha^2} \,.
\eeql
 $E^\nu ,T$ are time-dependent quantities, namely $T(t)=T_{now}/a(t)$ and
 $E^\nu(t)=E^\nu_{em}\frac{a(t_{em})}{a(t)}$, where $E^\nu_{em}$ is the 
\yy of a \zz at \emn.

 Thus after emission, the \zz has a mean-free-path 
$\lambda \sim a^5\sim t^{5/2}$:
\beql{lamt}
\lambda(t)=K \times  (t/t_{rad})^{5/2}\,, 
\eeql
where $t_{rad}$ is a constant chosen so as to give the appropriate value to
$a$ in the radiation-dominated epoch (see Appendix) .
One then has the length or time constant
\beal{k}
K\approx \frac{1}{\alpha^2}\frac{M^4}{T_{now}^4}\frac{1}{(E_{em}^\nu/GeV)
 a(t_{em})} \rm GeV^{-1}\\
\nonumber
\approx \nrnd{3}{38}\frac{1}{a(\txnd{em})(E_{em}^\nu/GeV)}\, \rm sec \\
\nonumber
\approx \nrnd{1}{48}\frac{1}{(\txnd{em}/sec)^{1/2}(E_{em}^\nu/GeV)}\, \rm sec
\eeal
with $E_{em}^\nu/\rm GeV$ equal to the \yy of a \zz at emission in $\rm GeV$, and
 $T_{now} =\nrnd{2.5}{-4}\rm eV$ the present \T of the CMB.
The probability of the \zzs from a \bu emitted at $t_{em}$ reaching a time
$t$ without interacting is thus
\beql{prob}
Prob(t,t_{em})\approx \exp\bigl[-\int_{t_{em}}^t\frac{1}{\lambda} dt
\,\bigr]
=\exp\bigl[-\frac{2}{3} \frac{t_{rad}}{K}\bigl(\,(t_{rad}/t_{em})^{3/2}
-(t_{rad}/t)^{3/2}.
\bigr)\bigr]
\eeql
With  $t_{rad}= \nrnd{1.9}{19} \rm sec$, the characteristic
dimensionless parameter  ${t_{rad}}/{K}$ in \eq{prob}   has the value, for say
 $\txnd{em}= 1 \rm sec, E^\nu_{em}=1 
 \rm GeV$, of   $\sim \nrnd{1}{-29} $.

 It is to be noted that in \eq{prob}
the \t $t$ appears in the exponent inversely and to a fractional power. This is quite different from
usual ``optical depth'' expressions, where it appears linearly. This
difference originates in the \t dependence of the medium and the \yy
dependence of the \zz \csssn.

\subsection{The Parameter \tx{free}}

The analysis becomes more transparent if we introduce a \t
 $\txnd{free}$,  defined as 
\beql{deftf}
\txnd{free}=\bigl(\frac{2\txnd{rad}}{3K}\bigr)^{2/3}\txnd{rad}\, .
\eeql
The interpretation of $\txnd{free}$ is given at the end of this subsection.
With this definition, \eq{prob} becomes
\beal{proba}
Prob(t,t_{em})
=\exp\bigl[-(t_{free}/t_{em})^{3/2}+(t_{free}/t)^{3/2}\bigr]\,,
\eeal
which also can be written as
\beal{probaa}
Prob(t,t_{em})&=&\exp\bigl[-(t_{free}/t_{em})^{3/2}\bigr]\times
\exp\bigl[+(t_{free}/t)^{3/2}\bigr]\\
\nonumber
&=&\exp\bigl[-\kappa^{3/2}\bigr]\times
\exp\bigl[+(t_{free}/t)^{3/2}\bigr]
\eeal
where we introduce  the ratio $\kappa$
\beql{kp}
\kappa=t_{free}/t_{em}
\eeql
The \t dependence is given by the second factor in \eq{probaa}. Although
this factor is a  decreasing function,
one notes that the
 expression  does not tend to zero for $t\to \infty $.
 Rather it  goes to 1, not zero, for large $t$.

 While $Prob(t,t_{em})$  thus no longer decreases as  $t\to \infty $,
  this can happen in two ways. The first factor,
$\exp\bigl[-\kappa^{3/2}\bigr]$, can be of order one, or it can be very
small, close to  zero. 
When it is of order one, it means that the \zzs have ``escaped'' to
 late \tn s without interacting. When it is near zero
 it means that the \zzs have not reached late
\tn s, they have been ``contained''.
 There are thus two regimes of behavior, according to whether one has
 $  \txnd {em}> \txnd{free}$, that is $\kappa <1 $; or
$  \txnd {em}< \txnd{free}$, that is $\kappa >1 $.

{\bf Cases  $\txnd {em}> \txnd{free}$:}
 the emission is
  later than  \tx{free}. {  Since $\txnd{free}/t<1$ } always, 
 the exponent in the second factor of \eq{probaa}
is always small and the factor 
 varies little.
 This is the
justification for the label ``free'', the \zzs are moving freely.
This ``escape''  reflects the
fact that the dilution of the medium and the redshifting of the \yy has
dominated over  the absorption of the \zzsn .

{\bf Cases  $\txnd {em}< \txnd{free}$:}
 the emission
takes place  earlierer than  \tx{free}.
 Then the exponent of the
second factor of \eq{probaa} is large and
 the factor decreases rapidly until the \bu
is  totally absorped. One sees that the period of rapid decrease ends when
$t \approx \txnd{free}$. Hence in these cases \tx{free} is the \t by which
the \zz pulse is absorped.

The significance of \tx{free} is thus that it determines if a \zz
``escapes'' or not, and if it does not, the \t by which 
 it is ``contained''.  As will be seen in the Table below, the
transition around $\kappa\sim 1$ is rather abrupt and we can regard
$\kappa= 1$ as a dividing line between the two regimes { of
``escape'' or ``containment''}.

\subsection{Numerical Estimates}

For the quantitative evaluation of \tx{free}, we write
\beal{free}
\txnd{free}=\bigl(\frac{2\txnd{rad}}{3K}\bigr)^{2/3}\txnd{rad}=
\biggl(\frac{2\,(\txnd{em}/sec)^{1/2}(E_{em}^\nu/GeV) \txnd{rad} }
{3\times \nrnd{1}{48}}\biggr)^{2/3}\txnd{rad}\\
\nonumber
=  \nrnd{4}{-20} \biggl({(\txnd{em}/sec)^{1/2}(E_{em}^\nu/GeV) }
\biggr)^{2/3}\txnd{rad}\\
\nonumber
=\nrnd{8}{-1} \biggl({(\txnd{em}/sec)^{1/2}(E_{em}^\nu/GeV) }
\biggr)^{2/3} \rm sec
\eeal
This result may also be written as 
\beal{freea}
\kappa=\frac{\txnd{free}}{\txnd{em}}
=\nrnd{8}{-1} \biggl(\frac{E_{em}^\nu/GeV}{\txnd{em}/sec}
\biggr)^{2/3} 
\eeal

\begin{table}
\begin{center}
\begin{tabular}{|l|l|l|l|l|l|}
\hline
\tx{em}/sec&{ z} &$E^\nu_{em}/GeV$&$E^\nu_{cmb}/GeV$ &\tx{free}/sec&$Prob_\infty$\\
\hline
\hline
\tend{13}&\tend{3}&\tend{14}&\nrd{1.0}{14}&\nrd{3.6}{13}&\nrd{1.1}{-3} \\
\hline
\nrd{5}{12}&\nrd{6}{3}&\tend{13}&\nrd{7.4}{12}&\nrd{6.3}{12}&0.24 \\
\hline
\tend{12}&\nrd{4}{3}&\tend{17}&\nrd{3.3}{16}&\nrd{1.7}{15}&0.0 \\
\hline
\tend{12}&\nrd{4}{3} &\nrd{1}{3}&\nrd{3.3}{2}&\nrd{7.9}{5}&1.0 \\
\hline
\tend{12}&\nrd{4}{3} &\nrd{1}{10}&\nrd{3.3}{9}&3.7&1.0 \\
\hline
\tend{10}&\nrd{4}{4} &\tend{6}&\nrd{3.3}{4}&\nrd{7.9}{3}&1.0 \\
\hline
\nrd{9}{8}&\tend{5} &1&\nrd{9.9}{-3}&\nrd{7.6}{2}&1.0 \\
\hline
\nrd{9}{4}&\tend{7}  &100&\nrd{9.9}{-3}&\nrd{7.6}{2}&1.0 \\
\hline
\nrd{9}{2}&\tend{8} &1000&\nrd{9.9}{-3}&\nrd{7.6}{2}&0.46 \\
\hline
9&\tend{9} &\tend{4}&\nrd{9.9}{-3}&\nrd{7.6}{2}&0.0 \\
\hline
1&\nrd{4}{9}&\tend{5}&\nrd{3.3}{-2} &\nrd{1.7}{3}&0.0\\
\hline
0.1&\tend{10} &\tend{5} &\nrd{1.1}{-2}&\nrd{7.9}{2}&0.0 \\
\hline
0.1&\tend{10} &\tend{3} &\nrd{1.1}{-4}&36&0.0 \\
\hline
0.1&\tend{10} &\tend{1} &\nrd{1.1}{-6}&\nrd{3.7}{-1}&\nrd{8.1}{-4} \\
\hline
0.1& \tend{10}  &\tend{-1} &\nrd{1.1}{-8}&\nrd{7.9}{-2}&0.50 \\
\hline
\tend{-2}&\nrd{4}{10} &\tend{-1}&\nrd{3.3}{-9}&\nrd{3.7}{-2}&\nrd{2.3}{-10}\\
\hline
\tend{-2}&\nrd{4}{10} &\tend{-2}&\nrd{3.3}{-10}&\nrd{7.9}{-3}&0.50\\
\hline
\tend{-4}&\nrd{4}{11} &\tend{-3}&\nrd{3.3}{-12}&\nrd{3.6}{-4}&\nrd{1.1}{-3}\\
\hline
\tend{-5}&\tend{12} &\tend{-4}&\nrd{1.1}{-13} &\nrd{3.6}{-5}&\nrd{1.1}{-3}\\
\hline
\tend{-5}&\tend{12}  &\tend{-5}&\nrd{1.1}{-14}&\nrd{7.9}{-6}&0.50\\
\hline
\tend{-6}&\nrd{4}{12} &\tend{-6}&\nrd{3.3}{-16}&\nrd{7.9}{-7}&0.50\\
\hline
\tend{-9}&\tend{14}  &\tend{-8}&\nrd{1.1}{-19}&\nrd{3.7}{-9}&\nrd{8.1}{-4}\\
\hline
\tend{-10}&\nrd{4}{14}  &\tend{-8}&\nrd{3.3}{-20}&\nrd{1.7}{-9}&\nrd{3.6}{-31}\\
\hline
\end{tabular}
\end{center}
\caption{
Some  parameters for  the \zzs  given various  emission times
and  \yys at emission, using our simplified model. { The second column,$z$,
 is the redshift corresponding to
  the \em time,} and  the third is  the \em \yyn.
   The fourth column is the \yy after a \rsh
to  \tx{cmb} , using
  $E^\nu_{cmb}=\nrnd{3.3}{-7}(\txnd{em}/sec)^{1/2} E^\nu_{em}$, and the
fifth column is \tx{free}, as found from \eq{free}.
At an \em \t of 0.1 seconds a \zz with \yy below 100 MeV can escape,
while at \tend{-6} sec. the \yy must be below 1 keV.
 One notes that the \prob of an escape to large
times  $P_\infty $  becomes essentially zero when the
 emission time is before  \tx{free}.
In these cases, with $P_\infty \approx 0$, \tx{free} gives the time by which
 the \zzs  have interacted (see text).}
\label{tab}
\end{table}

From the form of \eq{freea}, one sees that a constant value of $\kappa$,
which gives the ``escape''  probability, arises from a constant ratio between
$E_{em}$ and \tx{em}. In particular 
$\kappa=1$ corresponds to the condition
\beql{k1}
 \txnd{em}/sec= \nrnd{7}{-1}\, E_{em}^\nu/GeV~~~~~~~~~~~~
x_{em}=\nrnd{8}{-14} E_{em}^\nu/GeV\,,
\eeql
where $x_{em}=t_{em}/\txnd{cmb}$ refers to
 a dimensionless scaled \t variable $x=t/t_{cmb}$  that we shall use below.  We shall refer to
the parameter $\nrnd{8}{-14}$ as ``$s$'' in the following:

\beql{sdef}
s=\nrnd{8}{-14} \, .
\eeql

These numerical values arise from our simplified assumptions, which have
allowed us to  identify the qualitative features of the problem.
While the numbers  may be expected to  
change in more detailed treatment, the general features may be expected to
remain, namely a division of the $(\txnd{em},E_{em}^\nu) $ space into
an ``escape'' and a ``confined'' region, with an approximately linear,
increasing, boundary between the two.

  Table 1  shows some of the parameters
  for sample
 emission time and emission \yyn, using \eq{free}.
 The first column is the
\em \tn, the second the corresponding redshift $z$, the third column
the  \yy of the \zzn, the fourth the \yy after the \rsh
to \tx{cmb}, the fifth $\txnd{free}$.  The last column is  the
quantity  $P_\infty$, the first factor in \eq{probaa}, giving the
\prob of ``escape''  to large \tn s:

\beql{inf}
Prob_\infty =\exp\bigl[-(t_{free}/t_{em})^{3/2}\bigr]
=\exp\bigl[-\kappa^{3/2}\bigr]\,,~~~~~~~~~~
\eeql
Since \tx{cmb} will generally be much later than our other
times, $Prob_\infty$  will  also be the \prob that a \zz reaches the
formation \t of the CMB
or the \t of the last \sctgn ,  \tx{cmb}. One sees how the \prob of ``escape''
essentially depends on whether the \em is before or after \tx{free}.

\vskip1cm

\begin{figure}
\begin{center}
 \includegraphics[width=1.1
 \linewidth, angle=0]{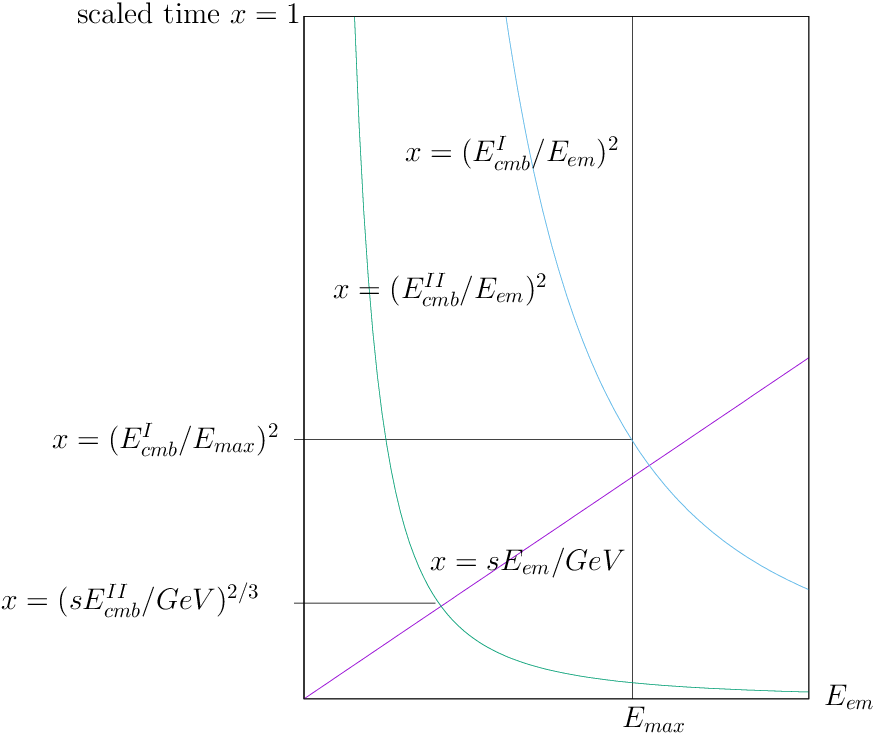}
 \caption{ The (\em \yyn, \em time) plane for the free flight of \zzsn. $E_{em}$
is the \em \yy and $x$ is the scaled \em \tn . The plot   
 shows which  regions  can or cannot
 contribute to a given $E^\nu_{cmb}$ at scaled \t  $x=1$. The inclined line
rising to the right, $x=s\,E_{em}/GeV$, separates the plane into the regions
of ``escape'' (above the line) and ``contained'' (below the line).
 In the well-defined $\eb$ model, the potentially 
 contributing regions would correspond to a vertical band at
 some $E_{em}=\eb $ (not shown). For the spread \sp model, where all \yys up
to a certain maximum $E_{max}$  can occur,
  the region to the left 
of $E_{max}$ potentially contributes. 
 Points that can contribute to a given  $E^\nu_{cmb}$ are found
 by following a curve down from the point where it starts at $x=1$. 
The two curves are examples for \pls with different \yys
 at $x=1$. The curve to the right, 
Curve I,  is for a relatively
higher \yy  and  the curve to the left, Curve II, is  for  a lower \yyn.
 Curve I encounters the $E_{max}$
limitation before encountering the `escape' limitation. For curve II
it is the other way around.
These two possibilities lead to the different lower limits of integration
in \eq{nudi}.}
\label{regions}
\end{center}
\end{figure}

\vskip-2cm
\subsection{The $(\t, E_{em}) $ Plane}
Visualization is aided by a sketch
of  the $(\t, E_{em}) $ plane, shown in Fig.\ref{regions}.  The figure
shows the separation in this plane into regions for   ``contained'' or ``escaping'' \zzs. 
 The \t is represented by  the vertical axis, but since we will
be mainly concerned with \zzs at \tx{cmb}, we use the dimensionless
scaled \tn,
\beql{xdef}
 x=t/\txnd{cmb}\, . 
\eeql
  The horizontal axis is the \em \yyn.

 The  $\kappa =1$ line   is shown ascending to the
right,  given by  $x= s  E_{em}/GeV$,  as
in  the second form of \eq{k1}.
This line  separates the plane into regions of
`escape' (above the line) and `contained'  (below the line),
  according to  whether $\kappa$ is greater than
or less than one. 

 It is also interesting to show where in the plane \zzs  which can arrive
to \tx{cmb} with a certain \yy   originate. 
The curves for a given $E_{cmb}^\nu$ are sketched using the radiation-dominated
$x=(E_{cmb}/E_{em})^2$ formula for \rshn.
Following a curve down from $x=1$ to earlier times or smaller $x$,
 gives the points which
potentially contribute to a given \yy of the \zz \sp
 at \tx{cmb}. However if the curve goes
below the  $x= s  E_{em}/GeV$ ``escape'  line, that region cannot contribute. Similarly
if the \em \sp does not extend beyond a certain $E_{max}$, the region to 
the right of $E_{max}$ is also excluded. A curve enters the forbidden region
for `escape' at \t $x=(s E_{em}/GeV)^{2/3}$ and when there is an
  $E_{max}$ constraint, at
\t $x=( E_{cmb}/E_{max})^2$. 

An interesting conclusion from these  ``escape'' considerations is that
a very low \yy \zz cannot originate from a very high \yy \zzn. The 
 curve for a small  $E_{cmb}^\nu $ will generally enter the forbidden zone
for `escape' before it gets to high \yyn. That is, the large
\rsh needed would imply \em at an early \tn, where it would however be stopped
before reaching later \tn s.

\section{Morphology}

A  \bu at \t \tx{em}  will spread out from its point of origin.
The furthest traveling component among the presently known elementary
\pls  and probably its principal one,
 will be the \zzsn. Since these travel essentially at the speed of
light, the  radius they reach at \tx{cmb} gives the outer radius of
 the region where possible signs of the \bus on the CMB will show.
 We say ``radius''
under the assumption that the \bu is spherical; if not our numerical
 results should be taken as  merely qualitative. 

We consider two situations, one where the \zzs (or other weakly interacting,
essentially massless, \pls) ``escape'', that is travel freely to  \tx{cmb}
and later, or
alternatively where they are absorped earlier. As explained
above, this will be given by the parameter \tx{free}, see section 2.2.

\subsection{Neutrino \bu Radius, ``Escape''}
With the \zzs  traveling  at the speed of light ( small
\zz mass and index of refraction effects are negligible) the 
radius or linear dimension for free flight until \tx{cmb} is given by
\beal{trava} 
a(\txnd{cmb})~ \int^{t_{cmb}}_{t_{em}}(1/a)dt
=
2\txnd{cmb}\bigl(1-\bigl(\frac{\txnd{em}}{\txnd{cmb}}\bigr)^{1/2}\bigr)
\approx \nrnd{2}{13} sec\\
\nonumber
 ~~~~~~~~~~~~~~~~~~~~~~~~\txnd{cmb}>>\txnd{em}
\eeal 
For early \bus, this is independent of emission time and
 corresponds to
 an angular dimension of \nrd{2}{-2} radians on the CMB.
In our symbolic representation Fig.\ref{blc}, we represent 
these ``escaped'' \bus by the larger little circles,
all of the same size.

\subsection{Neutrino \bu Radius, ``Contained''}\label{cntnd}

When the \zzs are ``contained'',  then the \bu \yy reaches out
  to a minimum radius  \tx{free}
 by ``free flight''. 
 That is, the \zzs travel freely  between the \tn s
\tx{em} and \tx{free}. A further propagation of the \bu might be relevant
by a diffusion-like process of the \pls produced by the \zz interactions.
These  will be essentially \pls with strong or electromagnetic interaction
with a much shorter interaction length, resulting in a diffusive behavior.
We give rough  estimates of the lengths involved. These lengths are relevant
both for understanding 
  the  possibly observable size of the resulting region on the CMB, and for
estimating the volume in which the \bu \yy is deposited.
 
\subsubsection{Free Flight}

The ``free-flight''  distance may be found as in \eq{trava}, with the difference that
the coordinate distance is given  by \tx{free} and not \tx{cmb}. The $a(t)$
factor, however, is still $a(t_{cmb})$ since we wish to obtain the proper
size at \tx{cmb}. Thus for contained \bus the radius or linear dimension 
at \tx{cmb} is
\beal{travb} 
  a(\txnd{cmb}) \int^{t_{free}}_{t_{em}}\frac{dt}{a}
= 2\txnd{cmb}^{1/2}\bigl(\txnd{free}^{1/2}-\txnd{em}^{1/2}\bigr)
\approx \nrnd{2}{13} \bigl(x_{free}^{1/2}-x_{em}^{1/2}\bigr)sec,
\eeal  
using the scaled \t $x=\frac{t}{\txnd{cmb}}$.

 As the Table shows, a great range of $x_{free}$ and $x_{em}$ are
 imaginable, so  unless the \bus are concentrated around
a certain \em \t with a definite \yyn , a large range of radii are possible,
up to $\sim \nrnd{2}{13} \rm sec$. On the other hand, should the 
\bus be mainly from a certain \t  $x_{em}$, then \eq{travb} could help to identify
that \tn .
In Fig.\ref{blc} we represent these ``contained'' \bus by the
 smaller little circles. These are 
of varying  size, in accordance with \eq{travb}.

\subsubsection{Diffusion Length}\label{dff}

A second mechanism for the spread of the
\bu   is possible,  namely diffusion-like processes.
When
the \zzs interact, most of the \yy will be transferred to an electromagnetic
or hadronic component, in reactions of the type 
\beql{ecomp}
\nu \to e^-....,  ~~~~~~~~~~~\nu \to \mu^-....
\eeql
where $...$ can be hadrons, quarks, or other components of the electroweak
standard model. These components, having electromagnetic or strong
interactions, will have a much shorter interaction length than \zzsn.
 We estimate a diffusion parameter
based on their \csssn s.
 The electromagnetic component
is characterized by a \csss $\sigma \sim \alpha^2/m_e^2$ and the strong
interaction component by $\sigma \sim  1/(100 MeV)^2$. Both of these  
are roughly  energy-independent and have  about the same size, so we treat them
together. This separation into  weak and other components is a simplification
since there will be transfers between the components, but after the \zzs
interact most of pulse will consist of the non-weakly interacting
components.

Diffusion may be viewed as a random walk process, and 
to find a diffusion length, 
we   use a random walk argument
with  the above electromagnetic/strong interaction parameters.

{ Because we have a small, isolated \sy we are able to treat the problem
in a simple way, without   genearal-relativistic considerations and
particular coordinate \sys. We use the Equivqlence Principle, which says
that in
a small ``freely falling'' \sy the \gvtl field is eliminated \cite{LL}.
      We thus  place ourselves in the rest frame around  the origin of the
\bun. There is no \gvtl field and   
the problem then becomes one of a random walk in a medium  of decreasing
density. The length found is then automatically a proper length.} 
In such problems, one finds the variance $Var$   in the probabilistic
  location of the
\pln, given by  the number of collisions. The square root
of the variance then gives the characteristic length.

The size of the step in the random walk here is given by the distance between
collisions or  
the  mean-free-path. Using  $\sigma \sim  1/(100 MeV)^2$ 
from $\lambda =1/(\rho \sigma)$ one has
\beql{mfpa}
\lambda =\frac{(100 MeV)^2}{T^3}
=\frac{(100 MeV)^2}{T_{now}^3}(t/t_{rad})^{3/2}
= \nrnd{4}{11} (t/ t_{rad})^{3/2} \rm sec
\eeql
At $t=1\, \rm sec$, for example, one has  $\lambda\sim \nrnd{6}{-18} \rm sec$, and
around \tx{cmb} some hundreds of seconds. At very earliest \tn s the interaction
lengths become very short and  one might doubt the applicability of naive
kinetic theory. However, as we shall see, we are mainly concerned with the
epoch around \tx{cmb} and
 we will continue to
use this simplest approach.

To find the distance  
 for the random walk,
 we estimate the variance $Var$
in its distance from its point of origin. For a
  particle at light velocity,  one has in a short time $dt$
\beql{l2}
 d(Var)= \lambda^2  dN=\lambda^2 \frac{ dt}{\lambda}=\lambda \, dt\,,
\eeql 
where dN is the number of collisions in the time $dt$.
With 
\eq{mfpa}, one then has for an interval up to $t_{cmb}$
\beal{vara}
Var=\int_{t_{em}}^{t_{cmb}}\lambda \, dt&=&4 \times \tennd{11} 
\int_{t_{em}}^{t_{cmb}}(t/t_{rad})^{3/2} dt\, sec\\
\nonumber
&=&8\times \tennd{30}
\biggl(\frac{t_{cmb}}{t_{rad}}\biggr)^{5/2}\biggl[1-
\bigl(\frac{t_{em}}{t_{cmb}}
\bigr)^{5/2}\biggr] sec^2 \,.
\eeal
For  $t_{cmb}>>t_{em}$, the last bracket may be  set to one. Thus
for all very early  \bus the diffusion distance $ \surd{Var} $ 
  at \tx{cmb}  is about the
same. It has the value roughly
\beql{dfd}
\surd{Var}=\sqrt{8\times \tennd{30}\biggl(\frac{t_{cmb}}{t_{rad}}\biggr)^{5/2}}
~~ sec\approx \nrnd{4}{7} \rm sec\,,
\eeql
or about a light year.

 Due to the short length for strong/electromagnetic interactions we arrive
 at a very  small distance, cosmologically speaking. This justifies our view 
of the problem as one for a ``line'' in spacetime, where the \gvtl field is
absent, and so we can  use non-general
relativistic arguments \cite{LL}. If we ask over what scale the
``line''picture might break down, the only dimesnional scale in the  cosmology
is the time $a/{\dot a}$, which at $t_{cmb}$ is about $10^{13}$ s. This is
much more than a year, so that the \sy is
 in fact  well represented by a ``line''.   
  In the following we shall take the size of a 
``diffusive spot'' at \tx{cmb}
 to be $d \times ly$, with $d$ a parameter of order one.

 The  size \eq{dfd} for the diffusive region is a
constant,  does not depend on the \tx{em}, while on the other hand the size
for the ``free flight'' region \eq{travb}  does   have this dependence.
The qualitative reason for this difference is that the ``free flight''
distance results from the flight \t and so involves the difference between
the \em \t and the absorption \tn. But in the random walk behaviour for the
diffusive spread, the rapidly increasing step size  as the ambient
density decreases means that the size will be determined only by the 
last  steps. This is seen in the behaviour of the integral \eq{vara}.

Which of the two mechanisms gives the size of a contained \bu
at \tx{cmb} will depend on a comparison of \eq{dfd}
 (or similar calculation \cite{segre} )
 with \eq{travb}. The results here suggests then  for \tx{free}
relatively early, before a time of seconds, that  the
``free flight'' size  can  become less than the diffusion size.

While the exact values of the parameters 
 can  be changed in a more detailed treatment, these qualitative features
 may be expected to remain. 

 Of course, by causality neither of these distances can be 
larger  than that for the simple non-interacting
   light-like flight to \tx{cmb}. Comparison 
with \eq{trava}  shows  this is indeed the case.

\section{ Some Quantities } \label{dfns}

For the different hypotheses,  a few fundamental quantities  
are needed,  which we summarize in this section. Since we will be principally concerned
with effects on the CMB, we often  use the scaled dimensionless time variable $x$:
 \beql{xdefa}
x= \frac{t}{\txnd{cmb}}= \nrnd{1.1}{-13}{t/sec}\,.
\eeql
Since we will almost always be dealing with very early times, we will
mostly have $0<x<1$.

\subsection{Redshift  to \tx{cmb}}

The \rsh of an \yy $E_x$  given  at scaled time $x$ to the \t 
 \tx{cmb} is, in view of the
$a\sim t^{1/2}$ behaviour of the cosmological  scale factor $a$,

\beql{escale}
E_{cmb}=x^{1/2} E_x.
\eeql

\subsection{Burst \yy}
 Our working assumption  is that the total \yy of a \bu at \em is the \yy within
 the horizon at that \t 
(see \cite{jl}, Eq 29) (the subscript `pl' refers to  ``planck''), namely 
\beql{bue}
E^{tot}=M_{pl}\frac{t}{t_{pl}}= x\,M_{pl}\frac{\txnd{cmb}}{t_{pl}}
= \nrnd{2}{56}\, x\,M_{pl}\, ,
\eeql
with $x$ the scaled \em \tn. 
At \tx{cmb} this \yy will be, using \eq{escale},
\beql{buea}
E^{tot}_{cmb}= x^{3/2}\,M_{pl}\frac{\txnd{cmb}}{t_{pl}}
= \nrnd{2}{56}\, x^{3/2}\,M_{pl}
\eeql
The effects we shall discuss are linear in the \bu \yyn, so if one assumes
a reduced fraction of the horizon \yy is more appropriate, our results can be
simply multiplied by this fraction.

\subsection{``Escape'' Condition} \label{escp}
The ``escape condition'' determines the earliest  \t a \zz
with a certain \yy can  be emitted
 and still propagate to late \tn s without being absorped. In \eq{k1}
there is a  linear relation between the \em \t and \em \yy which results
in $\kappa= \txnd{free}/\txnd{em}=1$.   Since there is a relatively
abrupt transition around $\kappa=1$ from ``escape'' to ``contained''
(see examples in the Table),
 we take $\kappa=1$  as defining when `escape'
occurs. In terms
of the scaled \em \t $x$, the condition to ``escape''  is then
\beql{borderx}
x>s \times  E_{em}^\nu /GeV\,~~~~~~~~~~~~~~~~s=8\times 10^{-14}.
\eeql
Or, given an \em \t $x$, the condition for a \zz to  ``escape'' is
\beql{bordery}
E^\nu_{em} < s^{-1}\,  x\, \, \rm GeV \,.
\eeql
We may also need these conditions expressed in terms of the \zz \yy
at \tx{cmb}. Using  $E_{em}^\nu =x^{-1/2} E_{cmb}^\nu $, the condition
\eq{borderx} becomes

\beql{min}
 x >(s\times  E_{cmb}^\nu /GeV )^{2/3}\,.
\eeql

\subsubsection{Very High Energy Neutrinos  }
Where the $x=s \times  E_{em}^\nu /GeV$ line crosses the $x=1$ line, 
(Fig \ref{regions}) determines
the highest \yy \pls  that could reach \tx{cmb}. We thus infer a maximum 
\bu  \zz  \yy possible
at \tx{cmb} 
\beql{emax}
  s^{-1} GeV \sim \tennd{15} \rm GeV.
\eeql
This is below the planck scale of $\sim \tennd{19} \rm GeV $
  but above the highest
\yy observed for  cosmic rays $\sim \tennd{12} \rm GeV$.

  Neutrinos with very high  energies  at present  are  searched for by arrays
 like  GRAND \cite{Kotera:2021hbp}, AUGER
\cite{Gonzalez:2022erx}, ICECUBE \cite{gaiser} and
KFb  KM3NeT \cite{km3} . So far the highest
\zz \yys discussed by these groups have been in the PeV (\tend{6} GeV)
or the multi-PeV range.

Such \yys at present could arise from even higher \yy \bus 
 at  early \tn s
 which are then \rshn ed to the present. However, because
of the ``escape'' restriction the \bus cannot be too early. One can examine
the question by using Fig.\ref{regions}. To use our simple formalism
with $a\sim~t^{1/2}$, it is best to transfer the \yy to appoximately \tx{cmb},
applying the suitable \rshn. Thus an \yy $E_{now}$ observed
at present could originate from an earlier (scaled) \t $x$ as given by a
decending curve on the plot by $x\approx (\tennd{3}E_{now}/E_{em})^2$. When this 
curve enters the  region below the slanted line. the \zz
cannot ``escape". According to the labels on the vertical axis of the
figure, the earlies time for the \em is then 
 $x= (s \, \tennd{3}E_{now}/GeV)^{2/3} .$ If one wishes to additionally
introduce an upper limit to the possible \em \yy such as the planck scale,
then this would be represented by the $E_{max}$ vertical
 line on the figure. One then
determines, for a given $E_{now}$, if the descending curve encounters the
 $E_{max}$  or ``escape" restriction first.

For an example, we consider the possibility that the \tend{8} GeV event
of \cite{km3} orginated at even higher \yy in the very early universe. We
assign it an \yy of  \nrd{1}{11} GeV at \tx{cmb}. It's possible \em \yy and
\em \t is found on Fig. \ref{regions} by following the downward curve
$x=\bigl(\nrnd{1}{11}/E_{em}\bigr)^2$ from $x=1$.
According to the labels on the vertical axis of the figure,
 this curve enters the
``forbidden to escape'' region at $ x=(s \,\nrnd{1}{11} )^{2/3}\sim
\nrnd{4}{-2}$. This  is then the earliest possible  (scaled) \em \tn, where
it would have had an \em \yy of $\sim \tennd{12}\, GeV.$

That high \yy cosmic rays might originate from violent events in the 
very early universe would be a new and intriguing avenue to the classic
question  of the origin of high \yy  cosmic rays.
 However, due to the ``escape question'' it would only 
seem to be relevant for \zzsn,---unless of course one considers `knock-on`"
effects from these \zzsn.

\subsection{ Burst Density}\label{budn}

We will generally calculate some quantity of interest by integrating over
\em \yys at a given \tn, and
then integrating over \tn. 
	In doing so, the conditions for `escape' must be
 respected.

For the \t integration
we  will need the density of the  \busn. This density will be diluted by the
expansion,  but  since we are mainly concerned with 
observations on the CMB, we give it as seen at \tx{cmb}.
In \cite{jl} we characterized the \prob  of a \bu
per unit 4-volume by the dimensionless $\cal P$ , normalized
 (Eq 8 of \cite{jl})
 such that the 3-density of \bus occuring at \t $t$
 in an interval of cosmic \t $dt$ is $\frac{3}{64 \pi t^4} {\cal P}(t) dt$,
This normalization
is such that $\cal P$$=1$ would imply that the \prob of a \bu in a time
interval $dt=t_{horizon}$ and in a horizon-sized spatial
 volume at \t t  would be 1. While this definition is tied to the FRW coordinate
\syn, its normalization to the horizon size, which is a physical quantity,
gives it a certain invariant meaning.  
Expressed in terms of $x$, one has, after scaling
the density as $a^3\sim x^{3/2}$, the contribution from a time interval
$dx$ to the \bu  number~density~ $\rho_{\bus}$   at~\tx{cmb} is

\beql{budensb}
d\rho_{\bus}= \frac{\nrnd{2}{-41}}{sec^3} 
 \frac{{\cal P}(x)}{x^{5/2}}dx\,
\eeql
where $x$ is  the scaled \em \tn.
 In assuming that  $\cal P$ depends only on the \t variable
 we make a homogeneity assumption.

 The definition of $\cal P$ was
motivated by the idea that the natural unit is the causal volume.
 The great number of casual horizon-sized regions
at early \tn s is reflected in the possibly singular behaviour
 of \eq{budensb}  near $x=0$.
 For example, a constant $\cal P$, with a cutoff
at the planck time of $\tennd{-43}\rm  sec$ would introduce a factor of
 $\tennd{84} $.

 Like the number of stars on a patch of the sky, the number of \bus
in a region under observation does not depend on  \tn . The question
 of possible  \t dependencies
is examined  in section \ref{tdep}.

\begin{figure}
\includegraphics[width=\linewidth]{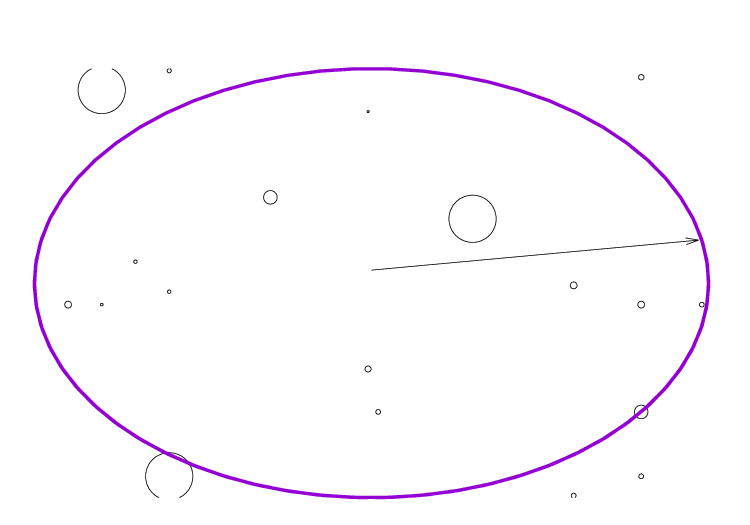}
\caption{
Symbolic, in only two spatial dimensions  and not to scale,
  representation of  how   \bus 
would appear on the last \sctg surface (lss)  at \tx{cmb}.  The large thick
 circle stands for the intersection
of our backward light cone (BLC) with the lss. Its radius is indicated by
the arrow and its thickness is meant to represent the finite size of the
observable region. The radius
increases by \tend{-3} ly in one earth year.
 The larger of the   smaller circles represent``escaping"  \zzsn,
 which reach  \tx{cmb} or later times. 
 According to \eq{trava}, these are all of about the same size.
 The smallest circles  represent ``absorped'' \busn, whose radius is
  governed either
by \tx{free}, or the ``diffusion distance''.}
 Observations via \phos are given  symbolically by  the intersection of the large circle
 with the \bu areas. In 2D this is a line segment, but in the  real
 3D case the circles become spheres, so the intersection
 with the BLC becomes a disc.
We  have assumed that the bursts are spherically symmetric.
\label{blc}
\end{figure}

A symbolic representation, in only two spatial dimensions and not to scale,
 of the situation at
\tx{cmb} is shown in Fig. \ref{blc}. The large circle represents our
backward light cone (BLC) at \tx{cmb}. This indicates
where  features on the CMB might be observed. The large circle has a
definite thickness to represent the spread in the observed region as given by
$d_{thom}$ the depth from which low \yy \phos can escape
in order to be detected (see \eq{h2}).
  The  smaller circles represent how
 the \bus could appear at \tx{cmb}. The larger of these represent
``escape"
 cases where the free flight of \zzs (or other essentially massless \pls)
is possible up to the \t \tx{cmb}. The smallest circles represent
"contained" cases where
this is not possible, that is, where the free flight distance is limited by
\tx{free}. In three spatial dimensions the circles become  spheres and so
the intersection of the BLC and the \bu regions is disc-like.

\subsubsection{Observational Depth}
For observations via low \yy \phosn, as for the CMB, there will be a certain
``depth" which can be observed at a given \tn . This is symbolized by
the thickness of the big circle in Fig.\ref{blc}.  We take this
distance $d_{thom}$  to be given by thomson \sctg
\beql{h2}
d_{thom}=\frac{1}{N_H\,\sigma_{thom}}=\nrnd{6}{21}cm=\nrnd{2}{11}sec \,,
\eeql
with ${N_H}\approx \nrnd{3}{2}/cm^3$ the density of hydrogen and  $\sigma_{thom}
=\nrnd{6.6}{-25}{cm^2}$ the
thomson  \csssn .

\subsubsection{Number of Bursts in an Observation}\label{buobs}
\eq{budensb} refers to the number of \bus per unit 3-volume.
For a given observational arrangement, the observed  3-volume
will be determined  by the angular acceptance of the observations, which lead
to a linear dimension on the CMB, $l_{acc}$, and the depth $d_{thom}$
 from which
\phos can escape, as conditioned by thomson \sctgn. Thus
the volume observed (by means of low \yy \phos) is
\beql{vobs}
V_{obs}=l^2_{acc}d_{thom}
\eeql
Hence the number of \bus  originating from a scaled
  \t interval  $dx$ is
\beql{nobs}
dn= V_{obs}\times\frac{\nrnd{2}{-41}}{sec^3} 
 \frac{{\cal P}(x)}{x^{5/2}}dx 
\eeql
For, say milliradian angular acceptance, one has $l_{acc}=\nrnd{9}{11} \rm sec$ 
 and with  $d_{thom}=\nrnd{2}{11} \rm sec$ one has 
  $V_{obs}=\nrnd{2}{35}\rm sec^3$.

\subsection{Burst Energy  Production}

We can use \eq{budensb} to express the total \yy  produced by the
\busn. Each \bu has an \yyn, according to \eq{bue},	
$E^{tot}= \nrnd{2}{56}\, x\,M_{pl}$, where $x$ is the scaled \em \tn.
. Since the \yy undergoes
a \rshn , it must be referred  to some \t or frame. We chose to express it 
in terms of the  value it would have at \tx{cmb}. 
According to \eq{escale} the \rsh  introduces a factor $x^{1/2}$, thus giving
a $x^{3/2}$ behavior. Then
combining with \eq{budensb} one has at  \tx{cmb}
the contribution to the \yy density $\epsilon_{\bus}$  coming from the \bus 
\beal{econtrb}
d\epsilon_{\bus}
 &&=\nrnd{2}{56} x^{3/2}\frac{M_{pl}}{sec^3}\times \nrnd{2}{-41}
 \frac{{\cal P}(x)}{x^{5/2}}dx \\
\nonumber
&&=\nrnd{3}{15}\frac{M_{pl}}{sec^3}\frac{{\cal P}(x)}{x}dx
\eeal

Integrating this will give the  \yy density from the \busn, expressed at
\tx{cmb}. It is noteworthy that the potentially strong singularity at $x=0$ has
becomes  only logarithmic  when considering the \yy density. This
conclusion  was anticipated in \cite{jl}, see  section III, `Olber's Paradox'.

\section{``Escape", Illustrative Models }\label{ill}

 As one sees from the Table or Fig.
\ref{regions},
\zzs which  are produced very early or 
with high \yy will generally interact before
\tx{cmb}, they will not ``escape''.
However, in this section we
 discuss how \busn, or parts of \busn , may indeed propagate
freely to $x=1$ or
\tx{cmb}.

The parameters involved, especially \tx{free},
 depend on the \pl \em \yyn.
 To examine the effects of this  \spn, we consider
 two simple models, one with a well defined \yy for the \bu \pls  and
one with a spread out \spn. We carry out the analysis using \zz parameters,
and estimate two quantities: the total \yy delivered to \tx{cmb} and the
\yy \sp at that \tn.
 To obtain these quantities at other times one may one may re-scale
densities or \yysn.. For 
 the present \t instead of at
\tx{cmb}, one  would use   
 $a(\txnd{cmb})^3=\nrnd{1}{-9}$ for density
 and rescale \yys by $a(\txnd{cmb})=
\nrnd{1}{-3}$. 

\subsection {General Formulas}

Before turning to the models, we note it is possible to give general formulas
in terms of the spectral \yy density at \emn.
That is,  since
a \bu can contain a range of \yys,  we introduce the quantity $\cal E$ for the \sp of the
\yyn. That is, the part of the  \yy in single
\bun, contained in an interval of \pl \yy at \em  $dE^\nu _{em}$  is:
\beql{cale}
 {\cal E}(E^\nu_{em}) dE^\nu _{em}
\eeql
With this definition, one has
\beql{inte}
 \int_0^\infty {\cal E}\, dE_{em}= E^{tot}\,,
\eeql
which is given by \eq{bue}. $ {\cal E}$  can also depend
on the \t variable, in general.
 ${\cal E}$ is dimensionless, being the ratio of two
\yysn.
If we prefer to think in terms of \pl number $N$ instead of spectral 
\yy density, one can write the pulse \yy as the product of the number
of \pls times their individual \yy 
 $dE^\nu_{em}{\cal E}= E^\nu_{em} dN,$
 so that
\beql{nrat}
\frac{dN}{dE^\nu_{em} }=\frac{1}{E^\nu_{em}}{\cal E}\,.
\eeql

\subsubsection{Energy Density}

The \yy reaching \tx{cmb} from one pulse at \em  \t $x$, taking those \yys that escape
from \eq{bordery}, and applying  the \rsh is
\beql{ep}
 x^{1/2}\int_0^{ul}{\cal E}\, dE_{em}= x^{1/2} {\cal I}(x)\, 
\eeql
calling the integral ${\cal I}$, which is an \yyn
\beql{epa}
{\cal I}=\int_0^{ul}{\cal E}\, dE_{em}.
\eeql
 The upper limit $ul$
can be the `escape' limit for that $x$, namely $x/s\, GeV $, or some limit on the
\em \yy $E_{max}$.
 $\cal I$   can depend on the \t variable $x$  both through the limit of 
integration and through  ${\cal E}$. 
 Now using \eq{budensb}, we have finally for the escaping  \bu  \yy
density $\epsilon^{esc}_{\bus}$  at \tx{cmb}
\beql{budensbz}
\epsilon^{esc}_{\bus}=\frac{\nrnd{2}{-41}}{sec^3} 
\int_0^1 dx \,\,  {\cal I}(x) \frac{{\cal P}(x)} {x^{2}}dx\, .
\eeql

\subsubsection{ Spectrum }

In addition to the total ``escape'' \yy density, another quantity of interest is
 the \sp of the \zzsn, how they are distributed with respect to \yyn.
 This
information is contained in ${\cal E}$, which must be \rshn ed to \tx{cmb}.
We note that both the numerator and the denominator in \eq{cale} \rsh the
same way, so the ratio
 is unchanged by the \rshn . We thus have, at \tx{cmb}, for a single pulse,
expressed in terms of $E^\nu_{cmb}$
the  \bu \yy contained in an interval $dE^\nu_{cmb}$
\beql{calea}
 {\cal E}(E^\nu_{em})\, dE^\nu_{cmb}
= {\cal E}(x^{-1/2}E^\nu_{cmb})\, dE^\nu_{cmb}  ,
\eeql
where $\cal E$ is still the \sp at \emn. 
Summing over all \tn s $x$ by means of \eq{budensb}, the integral is
from the earliest \t possible for the \yy in question to $x=1$.

 Then  the \yy density per unit \pl  \yy  $dE^\nu_{cmb}$  at \tx{cmb},
 as a function of \pl or \zz \yy   $E_{cmb}^\nu$  is
\beql{nudist}
\frac{d\epsilon^{esc}_{\bus}}{dE^\nu_{cmb}}=\frac{\nrnd{2}{-41}}{sec^3}
\int^1_{ x_{min}}{\cal E}(x^{-1/2}E^\nu_{cmb})
 \frac{{\cal P}(x)}{x^{5/2}}dx \,,
\eeql
where $x_{min}$ is the earliest \t from which the $E^\nu_{cmb}$ under
consideration  can
originate. When this is given by the `escape' condition then
 $x_{min}=(s\times E_{cmb}/GeV)^{2/3}$. Or if it is given by an $E_{max}$ limitation,
 $x_{min}=(E_{emb}/E_{max})^{2}$.  Which limit applies 
depends on the $E^\nu_{cmb}$ under
consideration   (see Fig \ref{regions}.)

\subsection{ Models}\label{smod}

From the above general formulas, the total  \yy density
or the \yy \sp can  be evaluated from   models or information
on $\cal P, I$. The deduction of the \zz spectrum associated
 with a \bu is undoubtedly
a complicated matter, involving both the primary process and the
characteristics of the ambient medium. However, to exemplify 
the role of the various factors involved,  we will consider some
very simple---probably over-simplified--  models. In the first model
we assume that the \zz \em is concentrated around a well defined
\em \yy $\eb$. In a second model, we take the opposite case, where
 the \zz \yy at \em is uniformly spread out,  up to some $E_{max}$.
{Finally, we mention a third case where the \em all
 takes place at essentially a given cosmic \tn.}

\subsubsection{Well Defined  Emission  Energy}  \label{welldf}
A situation which can be examined in a straightforward manner is
that in which  the 
\zz \yy at emission has a relatively well defined value, which we call
$\eb$. While  the \zz pulse will of course have a spread of \yys 
at emission, we assume in this model that this spread is not large, compared
to  the average \pl \yy in the pulse $\eb$. We will  then
approximate the situation with a unique \yy for the \zzs at emission, 
with value $\eb$.

In this approximation the  integral over $E^\nu_{em}$ may simply be replaced by
$E^{tot}$, \eq{inte}. However, in figure \ref{regions}
 below  where the exclusion line $x=s E_{em}^\nu/GeV$
would cross an $\eb$ vertical band, the \zzs cannot escape. 
 Thus for $\cal I$  we simply have
\beal{cli}
{\cal I}=E^{tot}_{em}~~~~~~~~~~~~~~~~x>s\times \eb /GeV \\ 
\nonumber
{\cal I}=0~~~~~~~~~~~~~~~~x<s\times \eb /GeV\, .
\eeal

{\bf Energy density :} With $E^{tot}$
 given by \eq{bue}, the integral for the total \yy
density to \tx{cmb} is given by  \eq{budensbz} and
one has for the \bu \yy density escaping to \tx{cmb}
\beal{budensbx}
 \frac{\nrnd{2}{-41}}{sec^3} 
\int_{s\times \eb /GeV }^1 dx \,\,  E^{tot}_{em} \frac{{\cal P}(x)} {x^{2}} \\
\nonumber
=\frac{\nrnd{3}{15}}{sec^3} M_{pl} \int_{s\times \eb /GeV}^1 dx
 \,\,   \frac{{\cal P}(x)}{x} \\
\nonumber
=\frac{\nrnd{4}{34}}{sec^3} GeV\times \int_{s\times \eb /GeV}^1 dx
 \,\,   \frac{{\cal P}(x)}{x}. 
 \eeal

{ \bf $\cal P$,$ \eb$  constant:}.\ If one were to
make this further drastic simplification
 that $\cal P$ and $\eb$ are constant in \tn , one would get for 
the \bu \yy density at \tx{cmb}
\beql{budensbu} 
\epsilon^{esc}_{\bus}=
\frac{\nrnd{4}{34}}{sec^3} GeV\, { \cal P}\bigl(ln(1/s) - ln(\eb/GeV \bigr) \,. 
\eeql
This can be compared with the  standard
 $(\pi^2/15)T^4=\nrnd{8}{33}GeV/sec^3$
for the  radiant \yy density at \tx{cmb}. Thus if  ${ \cal P}$ is relatively
large, this \yy could be significant.  However we stress that a basic
assumption of our calculations is that $\cal P$ is sufficietly small that
the standard cosmology is essentially unaffected. For some discussion
of this see the beginning of section \ref{ass}.

{\bf Neutrino  \sp:}
With the simplification  in this
model of a definite \yy $\eb $, we can also evaluate
the form of the \zz \yy \sp at \tx{cmb} from \eq{nudist}.
Approximating the \yy \sp density by a delta-function at $\eb$, one has
${\cal E}= E^{tot}\delta({\eb -E_{em}})$ so that
 \eq{nudist} for the total \yy density  per  interval of  $dE^\nu_{cmb}$ 
becomes 
\beql{nudista}
\frac{d\epsilon^{esc}_{\bus}}{dE^\nu_{cmb}}  =\frac{\nrnd{3}{15}}{sec^3}
M_{pl}\int^1_{ x_{min}}\delta(\eb-x^{-1/2}E^\nu_{cmb})
 \frac{{\cal P}(x)}{x^{3/2}}dx.
\eeql
 Taking into account the jacobian for
$  x^{-1/2}E^\nu_{cmb}$ in the delta -function, which is $2x^{3/2}/\ecmb$.
one obtains 
\beql{nudistb}
\frac{d\epsilon^{esc}_{\bus}}{dE^\nu_{cmb}}  =
\frac{\nrnd{6}{15}}{sec^3}
\frac{M_{pl}}{E^\nu_{cmb}} {\cal P}  \bigl( 1- x_{min}\bigr)
\eeql
with $x_{min}=(s\times  \eb /GeV )^{2/3}$ as given by \eq{min}.
For  $x_{min}>1$ the expression is to be taken as zero.
${\cal P}$ is to be  evaluated at the scaled \t $x=(E^\nu_{cmb}/\eb)^2$.
Using \eq{nrat},  this may be also converted to a \pl number \sp :
\beql{nudistc}
\frac{d N^{esc}}{dE^\nu_{cmb}}=\frac{\nrnd{6}{15}}{sec^3}
\frac{M_{pl}}{(E^\nu_{cmb})^2} 
 \bigl( 1- x_{min}\bigr) {\cal P} ,
\eeql
where $N^{esc}$ is the number density of ``escaped''  \plsn.
For a not too rapidly varying  $\cal P$, a 
$ E^{-2} $ behaviour
is  expected. We note this is quite different from a thermal
distribution, where there is  an exponential cutoff with \T .

There is of course a high \yy  cutoff in this model, for $E^\nu_{cmb}>\eb$,
since the \rsh of $\eb$ can only lead to lower \yysn.
However there is also a low \yy cutoff since there is a smallest
$x$ that can contribute, namely $x_{min}$. From the \rsh this lowest
$E^\nu_{cmb}$ is  $ x_{min}^{1/2}\eb =\eb(s\eb/GeV)^{1/2}$. On Fig,\ref{regions} this
corresponds to a situation where a vertical $\eb$ band would  cross a  curve 
at a point below the $x=sE_{em}$ line. 
One thus finds, in this model, that the narrow
\yy  band around $\eb$ at \em becomes, through the spread in \em \tn s, the band 
\beql{band}
 \eb(s\,\eb/GeV)^{1/2}< E^\nu_{cmb}<\eb
\eeql
at \tx{cmb}.
When we have $(s\,\eb/GeV) >1$, then all \em possibilities are ``under
water'', below  the suppression line $x=sE_{em}/GeV$
 on Fig.\ref{regions}. Since $s$
is a  small  number, (\eq{borderx}), it seems that a
wide range of $\eb$ are possible within this model. 
 When the \em \sp contains
\zzs down to arbitrarily low \yyn, as discussed for the 
`spread out' model below, a lower limit as in \eq{band}.
does not occur.

\subsubsection{Large Energy Spread Model  }\label{sprd}
In the previous section \ref{welldf}, we made the simplifying assumption that
that there was a well 
defined common \yy for the \zzs in a \bun.
 Here we  examine the opposite case, with a very spread-out \zz distribution,
reaching down to zero \yyn.
In this model unlike in the first model,
 there is no earliest escape  \t
since the \bus contain arbitrarily low \yy \zzsn.  
Also lower \yy and thus ``escaping'' \pls
need not come from higher \rshn, as was the case before.

As an example, we assume that at \em the total \yy in each interval of \pl
\yy is constant, up to some $ E_{max}$:
\beal{brems}
{\cal E} = const= \frac{E^{tot}}{ E_{max}}~~~~~~~~~~~~~~~~~~~~~~E^\nu_{em} < E_{max} \\
\nonumber
{\cal E} = 0~~~~~~~~~~~~~~~~~~~~~~~~~~~~~~~~~~~~~~E^\nu_{em} > E_{max}.
\eeal
This is like the spectrum in the   bremsstrahlung of \phosn, perhaps the ``most
spread out'' of familiar \pl spectra. The value of $const$ follows from the
fact that the integral over ${\cal E}$ should be
 $E^{tot}=  x\, M_{pl}\nrnd{2}{56}$.

To find the total \bu  \yy to \tx{cmb} we note that
 in this model all the \zzs in a \bu will ``escape'' if $ x> s \,
E_{max}/GeV$ (see Fig \ref{regions})
\beql{mxes}
{\cal I} = E^{tot} ~~~~~~~~~~ x> s \, E_{max}/GeV         
\eeql
and otherwise  only the part up to the escape \yy enters
\beql{mxesa}
{\cal I} = E^{tot}\frac{(x/s) GeV}{ E_{max}} ~~~~~~~~~~ x< s \, E_{max}/GeV.         
\eeql
Inserting in \eq{budensbz} gives the total escaped \bu \yy density at \tx{cmb}
\beal{budensby}
&&\frac{\nrnd{2}{-41}}{sec^3}
\int_0^1 dx \,\,  {\cal I}(x) \frac{{\cal P}(x)} {x^{2}}dx\\
\nonumber
&=&\frac{\nrnd{3}{15}}{sec^3}M_{pl}
\biggl(\int_{s \, E_{max}/GeV}^1   \frac{{\cal P}(x)} {x}dx
+\frac{1}{s E_{max}/GeV}\int^{s \, E_{max}/GeV}_o   {\cal P}(x)dx \biggr).
\eeal
This formula applies for $s E_{max}/GeV <1$, meaning there is some \t
before $x=1$ when the $E_{max}$ \zzs escape, when the full \zz \bu
escapes. When this is not true, when for all times $x<1$, only a part
of the \bu escapes,
then the first term is absent and we simply have
 the total escaped \bu \yy density at \tx{cmb}
\beal{budensbyy}
  \epsilon^{esc}_{\bus} = \frac{\nrnd{3}{15}}{sec^3}M_{pl}
\frac{1}{s E_{max}/GeV}\int^1_o   {\cal P}(x)dx 
\eeal
Since $s$ is small, which condition would apply depends essentially
on $ E_{max}$.

{\bf Energy with constant $\cal P$: }
Assuming constant $\cal P$ for illustration,
 one can evaluate \eq{budensby} explicitly to get for the 
\bu \yy density at \tx{cmb}
\beql{budensbyz}
 \epsilon^{esc}_{\bus}=
\frac{\nrnd{3}{15}}{sec^3}M_{pl} {\cal P}
\biggl( ln(1/(s E_{max}/GeV)) +1 \biggr) 
\eeql
where again with  $s E_{max}/GeV >1$ the log term is absent
and one has for the escaped \bu \yy density at \tx{cmb}
\beql{budensbyzz}
 \epsilon^{esc}_{\bus}=
\frac{\nrnd{3}{15}}{sec^3}M_{pl} {\cal P}
\frac{1}{s E_{max}/GeV}\,, 
\eeql
showing how a large  $E_{max}$  or spread reduces the `escape' .

In this model we can  give the \yy \sp using \eq{nudist}
so the \yy density per unit $dE^\nu_{cmb}$ is
\beql{nudistax}
\frac{d\epsilon^{esc}_{\bus}}{dE^\nu_{cmb}}= \frac{\nrnd{2}{-41}}{sec^3}
\int^1_{ x_{min}}{\cal E}(x^{-1/2}E^\nu_{cmb})
 \frac{{\cal P}(x)}{x^{5/2}}dx
\eeql
where $ x_{min}$ is the earliest \t  $E^\nu_{cmb}$, in view of
the escape condition,  can come from, namely
$x_{min}= (s\,E^\nu_{cmb}/GeV)^{2/3} $.
In  the present  model we have for ${\cal E}$ for a given
 $ E^\nu_{cmb}$  as a function of $x$

\beal{elim}
  {\cal E} &=& \frac{E^{tot}}{E_{max}}~~~~~~~~~~~x^{1/2}>
\frac{E^{cmb}}{E_{max}}\\
\nonumber
{\cal E} &=& 0~~~~~~~~~~~~~~~~~~~x^{1/2}<
\frac{E^{cmb}}{E_{max}}
\eeal

There are then two possible lower limits, $ll$,   on the $x$ integration;
either  $ x_{min}$ or the second of \eq{elim}, according to which is
larger. 
Then inserting $\cal E$ gives

\beql{nudi}
\frac{d\epsilon^{esc}_{\bus}}{dE^\nu_{cmb}}=
\frac{\nrnd{3}{15}}{sec^3}
\frac{ M_{pl}}{E_{max}}
  \int^1_{ll}
 \frac{{\cal P}(x)}{x^{3/2}}dx\,,
\eeql
where $ll$ is the larger of $(x\,E^\nu_{cmb}/GeV)^{2/3} $ or
$(E^\nu_{cmb}/E_{max})^2$ . It can be seen that two become equal 
where the escape line crosses the $E_{max}$ line, that is for
$x=s\,E_{max}/GeV$, or in terms of $E^\nu_{cmb}$, for
 $E^\nu_{cmb}=E_{max}(s\,E_{max}/GeV)^{1/2}$.
The situation is sketched in Fig.\ref{regions}.

{\bf Constant ${\cal P}$:}
Making the assumption of constant ${\cal P}$ for illustration, one then has  
for the total \yy density in $dE^\nu_{cmb}$
\beal{nudib}
\frac{d\epsilon^{esc}_{\bus}}{dE^\nu_{cmb}}=
\frac{\nrnd{6}{15}}{sec^3}
\frac{ M_{pl}}{E_{max}}{\cal P} 
~~~~~~~~~~~~~~~~~~~~~~~~~~~~~~~~~~~~~~~~~~~~~~~~~~~~~~~~~~~~~~~~~~~~~~~~~~~~\\
\nonumber
\times \bigl(\frac{E_{max}}{E^\nu_{cmb}}-1\bigr)
~~~~~~~~~~~~~~~~~~~~~~~~~~~~~~~~~~~~~~~E^\nu_{cmb}>E_{max}\times (s E_{max}/GeV) ^{1/2}\\
\nonumber 
\times \bigl( (s\,E^\nu_{cmb}/GeV)^{-1/3} -1  \bigr)
 ~~~~~~~~~~~~~~~~~~~~~~~E^\nu_{cmb}<E_{max}\times (s E_{max}/GeV) ^{1/2}
\eeal
With the value $\sim \tennd{-15}$  for $s$, the first option
falls away for $E_{max} > \tennd{15}\, GeV$ and the second variant
must be used for all \yys $E^\nu_{cmb}$ .
For very large $E_{max}/E^\nu_{cmb}$,
 which seems plausible for this model and for cases
where the first option still  applies, we can further simplify  to obtain

\beql{nudiba}
\frac{d\epsilon^{esc}_{\bus}}{dE^\nu_{cmb}}=
\frac{\nrnd{6}{15}}{sec^3}
{\cal P} \frac{ M_{pl}}{E^\nu_{cmb}}, 
~~~~~~~~~~~~~~~~~~~~~~~~~~~~~~~~~~~~E^\nu_{cmb}>E_{max}\times  (s E_{max}/GeV).
\eeql
 All these formulas may be exprssed in terms of a \pl number density
by $~E^\nu_{cmb}\to (E^\nu_{cmb})^2$ as in \eq{nrat}.

\subsection{Low Energy Early Burst Neutrinos at Present}\label{eab}

While many assumptions and approximations were made, a few conclusions
are suggested from the models. One is that, in general, a low \yy peaking
of the \bu \zz \sp  is to be anticipated. This 
reflects the higher \prob for a low \yy \zz to ``escape'',
combined with  the growing number of potential \em regions  at early \tn s,
\eq{budensb}, where the   \zzs   undergo a 
 very large \rshn .

Such  \zzs  from very early \tn s will become very low \yy \pls at present.
  With some  \zz mass, some or all of them
will likely  be  non-relativistic. As one sees from
the third column of the Table, it is quite possible that the very
 early \bu \zzs become non-relativistic, even if they have only
  millivolt massses.
It therefore appears that a prediction of the \bu model with \zzsn, as we have
discussed in this section, is that there is at present  a non-thermal population  
of low \yy \zzsn, with some or all of them non-relativistic.
These would be in addition to  the thermal $\sim 2\, Kelvin$
 \zzs expected \cite{nutemp}  in
any case from the standard cosmology.

In addition, in some of our estimates, as in \eq{nudistc}, there is a
power law  $\sim 1/E^2$
behavior of the \zz \spn. This is quite different from the exponential
cutoff in a thermal \spn. 

The difficult question of the detection of relic \zzs has a long and
interesting history \cite{wbg},\cite{spec},\cite{pt}. 

\subsection{Definite Emission Time Model}\label{tfx}
  A different type of simple model that might be used to illustrate some 
features of the question is one where the \bus essentially all originate
around the same  cosmic \tn. That is, we can 
 approximate ${\cal P}(t)\sim
\delta(t-t_{em})$.

 The most important aspect of such a situation 
would be the value of  the \em \t \tx{em}. This \t will determine the \rsh
to \tx{cmb} so that \yys will be shifted  according to \eq{escale}. If \tx{em}
is very early,  $x$ is very small and \pl \yys at \tx{cmb} are very small.
For example in the model with the given \em \yy 
$\eb,$ one will have
\beql{shit}
E_{cmb}=x^{1/2} \eb \,,
\eeql
where $x$ is (\tx{em}/\tx{cmb}) (\eq{xdefa}). If this down-shift is large
enough, say to less than an MeV, the \zzs will not be able to interact
inelastically and we would arrive to the situation mentioned
in section \ref{eab} where they escape to the present, forming a non-thermal
relic \zz background. This would also mean that the positron mechanism,
 to be discussed in section \ref{epl}, is not possible. With a spread \sp
this reasoning will apply to that part of the \sp which is down-shifted
below the inelastic threshold.

An important change vis-a-vis  the previous discussions
would concern \tx{free}. With \tx{em} fixed,  the only variability 
in \eq{free} is in the \yy factor. Thus if all the \bus are essentially similar,
\tx{free} would be the same for all \busn. Thus the smallest circles in
Fig \ref{blc} would all be of about the same size, and not variable as in
situations with a distribution  of \em times. By the same token,
approximately fixed values for \tx{free}and \tx{em} for all \bus would
mean that $\kappa$, governing  ``escape'' or
 ``confinement'',  is the same for all of them.
 
These considerations may be visualized on Fig \ref{regions}. Draw a
horizontal line at the scaled \t for \tx{em}. Along this line denote the
\em energy (energies). If the resulting point(s) are below the slanted line
the \pls do not ``escape''. If they are above the slanted line, construct
  curves like I or II starting  from the points. Where the curves
  cross the $x=1$ ordinate
gives their \yy at \tx{cmb}.

Concerning intensities in this model, let us express the $\delta$ function
hypothesis  in terms of the scaled \em time $x_{em},$ so that 
${\cal P} \approx {\cal P}_o\, \delta(x-x_{em})$,  where ${\cal P}_o$ is a 
dimensionless magnitude factor. Then according to \eq{budensb} we have a
number density of bursts at \tx{cmb}

\beql{budensba}
\rho_{\bus}= \frac{\nrnd{2}{-41}}{sec^3} 
 \frac{{\cal P}_o}{ x_{em}^{5/2}} \, .
\eeql
As an example, we could  apply this to the situation just mentioned,
 where the \zzs are below the inelastic threshold and escape to the present
to form part of the relic \zz background.  In the situation with a more or less
definite \yy at \em $\eb,$ one has for the number of \zzs in a \bu according to
\eq{bue},
\beql{bue2}
\nrnd{2}{56}\, x_{em}\frac{M_{pl}}{\bar E^\nu} \,.
\eeql
Multiplying by \eq{budensba} and a factor \tend{-9} for the \rsh of the
density to the present, one has finally

\beql{bue3}
N^{now}=\frac{\nrnd{4}{6}}{sec^3}\frac{M_{pl}}{\bar E^\nu} \,.
 \frac{{\cal P}_o}{ x_{em}^{3/2}}
\eeql
for the present density of relic \bu \zzsn,
in this model.

\section{Positrons }\label{epl}
If  \zzs  in the MeV range or above
  arrive at relatively late \tn s like \tx{cmb}
or afterwards, the production of positrons,
 whose annihilation could give 
  an observable signal, is possible.

The \dtn of such a signal would be very 
 characteristic of the \busn, since with
only $\sim eV$ or less  thermal  \yys present in the purely equilibrium
picture,  positrons should be absent.
That is we consider 
\beql{nuint}
\overline{\nu} +p \to e^+ + n~~~~~~~~~~~~~e^++e^-\to \gamma + \gamma 
\eeql
 giving  511 keV \phosn. Redshifted to the present \t these should give
soft  x-rays.  Below we atempt a more detailed estimate
 of the \rsh of the  annihilation \phosn.

 Other reactions which lead to
positrons via decay chains, such as $\overline{\nu_\mu}\to \mu^+ $ or
production of $\pi^+$ mesons could also contribute but \eq{nuint} has the
lowest threshold and should be the most significant.
  In principle there is also a 2.2 MeV $\gamma$
signal from capture
of the neutron on hydrogen, but due to the low matter
 density at \tx{cmb} or later the 14
minute decay of the neutron is much faster.

\subsection{Conversion Probability}
While of course positrons will be produced all along the path of a
\zz pulse, only those conversions  which are close enough so
that that the  annihilation \pho can ``escape'' to the present time
are potentially observable.

 It is possible to give
an estimate for the \prob of such a conversion without entering into
details of the intervening matter distribution.
   This is because the same atoms or matter density 
responsible for stopping the outgoing annihilation \phos are also the
relevant targets for the production reaction \eq{nuint}.
 Only when \eq{nuint} takes place within the
last mean-free-path for the  \pho will the \phos ``escape''. If $\tau$ is
the column density for this last mean-free-path, it is given by the
condition $\tau \sigma_\gamma(500keV)\approx 1 $ , or
 $\tau \approx 1/\sigma_\gamma(500keV) $. On the other hand the \prob for
a positron to be produced in this region is $\tau \sigma_{\nu\to e^+}$. 
Thus $\tau$ cancels and one is simply left with the ratio of the \csssn s.

 Thus the conversion \prob $C$  for a \zz to lead to an observable \pho  is
expected to be
\beal{convprob}
C=\frac{\sigma_{\nu\to e^+}}{\sigma_\gamma(500keV)}
=\frac{\nrnd{1}{-38}(E^\nu/GeV)\rm cm^2}{\nrnd{2.8}{-25}\rm cm^2}\\
\nonumber
\approx \nrnd{3}{-14} (E^\nu/GeV) \,,
\eeal
where we used  
$\sigma_{\nu\to e^..}=\nrnd{1.0}{-38}
(E^\nu/GeV)\rm cm^2$, with $E^\nu$ the \yy of the \zzs at the 
conversion \tn . For ${\sigma_\gamma(500keV)} $ see \eq{th}.

\subsection{Redshift of the Photon }

While the above argument shows that for an estimate of the conversion
\prob it is not necessary to know the exact locus where \eq{nuint} takes
place, we would also like 
to know  the \yy of the \anph at present. For this  it is necessary
to determine the epoch where the positrons  are produced to find the \rsh
of the 511 keV \phosn.

The quantity of interest is $N_\gamma(\txnd{now})$, the present density
of \anphn s, which  gives our \dtn rate. In particular we are interested
in  its \yy  \sp
$\frac{dN_\gamma(\txnd{now})}{d\omega}$,
where $\omega$ is the present \yy of an \anphn.

Although the annihilation of a positron on a stationary electron
leads to a ``line"  at 511 keV, various effects lead to a spread
of the \pho \spn. There is the thermal motion of the atoms and the
motion of the  bound electron in the atom. However, the most important effect 
in our present problem will be due to the spread of the \rshn s
in the origin of the gamma ray. There is an essentially one-to-one
relation between the \rsh of origin $z$ and $\omega$
\beql{oto}
\omega = (1/z) \,  511 keV ~~~~~~~~~~~~~~~~~~~~~~
~~~~~d\omega = (2/3) z^{1/2} \frac{dt}{\txnd{now}} \,  511 keV.
\eeql
Since we will find that the relevant times are well into 
   the matter dominated epoch, for z we use  
$z=(\txnd{now}/t)^{2/3}$, with $\txnd{now}=\nrnd{2.8}{17} seconds$ \cite{jl}. 
The second relation shows how a spread in production \tn s $dt$ gives
a band of present \yys $d \omega$.

A \t interval $dt$ at \rsh $z$  thus gives a contribution to the present
 density of the \anphn s with \yy $\omega = (1/z) \,  511 keV$
\beql{pf}
dN_\gamma(\txnd{now})= K(z) dt= K(z) \frac{1}{z^{1/2}}\,(3/2) 
 \frac{\txnd{now}}{511}
 \,  {d\omega} \,
\eeql
where $K$ is the   factor for the conversion of \zzs into \phosn,
involving   the local fluxes, \csssn s, and propagation to the present.

\subsubsection{Absorption Factor}

The most important element  in $K$ is  the
attenuation factor for the initially  511 keV \phon. That is,
  the \prob of an emitted  \pho to reach us is
governed  by a factor $\cal A$
\beql{cala}
{\cal A} =exp(-\tau/\tau_o)
\eeql
 where $\tau$ is the  column density or 
``thickness''   of
 the matter traveresed,   and 
$\tau_o $ is a parameter characterizing
 the matteer.
 This parameter is usually given in  grams/cm$^2$ and   for 500 eV \phos
in hydrogen one has \cite{nist}
\beql{th}
\tau_o\approx 6 \,grams/cm^2 \, .
\eeql

By referring to the mass of the  hydrogen atom or proton
 instead of grams, \eq{th}
may also be expressed as an effective \csss namely $\sigma_\gamma(500
keV)=\nrnd{2.8}{-25}cm^2$, which we used in \eq{convprob}. The
mass absorption  parameter or
\csss is \yy dependent and so will vary somewhat  during the flight of the
\pho as its \yy is redshifted. However, most of the absorption will take place
close to the \em \t and we  simply use the values for absorption at  500 keV.

To evaluate \eq{cala} in the cosmological situation, we take the intervening
matter to be essentially hydrogen and estimate the column density from the
present back to a \t $t$ or equivalently to an $a(t)$ or $z=1/a$. 
One first needs the density, which we take to be 
\beql{roo}
\rho(t)= \rho_o \bigl(\frac{t_{now}}{t}\bigr)^2\,,
\eeql
where $\rho_o$  is the present density of hydrogen and  the density
 scales as $1/a^3$, using  $a=(t/t_{now})^{2/3}$.

One thus has for the column density
\beql{roa}
\tau=\int_{t_{now}}^t \rho(t)dt
=\rho_o t_{now}\bigl(\frac{t_{now}}{t}\bigr)=\rho_o
t_{now}\times a^{-3/2}
=\rho_o\,t_{now}\times z^{3/2}\,,
\eeql
and for the absorption factor expressesed  in terms of $z$

\beal{cala1}
{\cal A} =exp(-\frac{\rho_o\,t_{now}}{6\, grams/cm^2 }\, z^{3/2})
&=&exp(-\nrnd{7}{-4}\, z^{3/2})\\
\nonumber
&=&exp(- (z/z_o)^{3/2})
\eeal
where we introduce the quantity
 $z_o=(\frac{\rho_o\,t_{now}}{6\, grams/cm^2)})^{-2/3}\approx 130$, which
characterises the absorption distance in terms of $z$.
We have taken $\rho_o$ at 5\% of the critical density, namely
  $\rho_o=\nrnd{5}{-31} grams/cm^3$.

\subsubsection{Spectrum}
 In addition   to $\cal A$ which
favors   nearby reactions, there are
  countervailing
factors favoring higher \rsh for \eq{nuint}. These   include the
 higher densities and  the smaller downshift of the \zz \yyn . 

We anticipate that these various effects will lead to a cumulative power
$\sim z^p$, so that the distribution of origins $z$
for a  presently observable \anph is proportional to
\beql{propz}
z^p {\cal A} =z^p \times exp(- (z/z_o)^{3/2}) \, .
\eeql
  
This product of increasing and decreasing funtions  leads to 
 a  peak at
some $z$, namely at $z=(\frac{2}{3} p)^{2/3}\times z_o$.

As for the value of $p$, the increasing density of target protons
 gives a factor $z^3$. The incoming \zz flux will also increase $\sim z^3$
but this is cancelled by a factor $1/z^3$ for the produced \pho density,
by the \rsh to the present. This cancellation reflects the fact that \pho
and \zz densities \rsh in the same way.

   As for the \zz \yy factor, this will depend  on whether we are in
the fully relativistic regime $E> 1\, GeV$ where the \csss is linear with \yyn,
or at lower \yyn, where the behavior is closer to quadratic. 
 With the $1/z^{1/2}$  factor from \eq{pf}, this leads to $p\approx
3.5---4.5$
and so
\beql{zpk}
z_{peak}\approx (1.8 --- 2.1) z_o \approx (230--- 280)\,,
\eeql
implying that the original 511 keV \pho appears around $ 511 keV/z_{peak}$
 that is as a soft,
$\sim  (2\,or\, 3)\, keV$ range x-ray.  

 An important  question is the width to be expected  in the observed
 \pho \spn, since this can be significant in separating
the signal from backgrounds.
Substituting  $z=511keV/\omega$  for $z$ in \eq{propz}, one finds
 a rather
broad  distribution in $\omega$:
\beql{propza}
\frac{dN_\gamma(\txnd{now})}{d\omega}\sim
 (1/\omega)^p exp(-(3.9/\omega)^{3/2})
\eeql

 Due to the spread
in the \rsh of  origin the ``line'' has become a   broad
 and somewhat asymmetric ``bump''.

\subsection{Intensity}
Finally, one  would like to have an estimate of the intensity of the
signal, that is the expected number density
of the \anphn s at present, $N_\gamma(\txnd{now}) $.
This will of course depend  crucially on the nature and frequency of the
\busn, but to see the interplay of the various elements,
 we can examine some of our simple schematic models.

A simple estimate follow by using the arguments of the previous section to say
that the \zz to \pho conversion approximately takes place at a
definite \tn. which we call \tx{C}. 
According to \eq{zpk} this  would  be the \t
corresponding to $z\approx 200$. Then 
\beql{tc}
N_\gamma(\txnd{now})\approx N_\nu(\txnd{C})\times C
\times \biggl(\frac{a(\txnd{C})}{a(\txnd{now})}\biggr)^3
\eeql   
The first factor is the number  density of relevant \zzs at \tx{C}, $C$ the
conversion \prob \eq{convprob},
 and the last factor the dilution factor to the present.

\subsubsection{$N_\nu(\txnd{C})$}
By the density of relevant \zzs we  mean
 those of sufficient \yy to  produce positrons. 
We call this \yy $E_{th}.$

Since the \zzs are not significantly absorped between between \tx{cmb} and
\tx{C}, we may refer  the calculation to our reference point \tx{cmb} and
perform the necessary \rsh to \tx{C}. One thus obtains

\beql{toC}
N_\nu(\txnd{C})= \int_{5E_{th}}^\infty  \frac{d
N_\nu(\txnd{cmb})}{dE^\nu_{cmb}} dE^\nu_{cmb}
 \times
\biggl(\frac{a(\txnd{cmb})}{a(\txnd{C})}\biggr)^3 \,.
\eeql
 The ``5" in the lower limit of integration results from the fact that the
\rsh at \tx{C} is about  one fiftth that as at our reference point
 \tx{cmb}. Inserting in \eq{tc},

\beql{tca}
N_\gamma(\txnd{now})\approx \int_{5E_{th}}^\infty
 \frac{d N_\nu(\txnd{cmb})}{dE^\nu_{cmb}} dE^\nu_{cmb}  \times C
\times \biggl(\frac{a(\txnd{cmb})}{a(\txnd{now})}\biggr)^3
\eeql

The cancellation of the $a(\txnd{C})$
 factor reflects the fact  the \zz and \pho
densities \rsh in the same way. The  cancellation shows that in a more
detailed treatment were one performs an average over conversion \tn s
\tx{C},
the only significant difference will be in the resulting average over  $E$.

Using $C\approx \nrnd{5}{-14}(E^\nu/GeV)  $ and the 
\rsh of \tend{-3} for \tx{cmb},
\eq{tca} becomes finally
\beql{tcb}
N_\gamma(\txnd{now})\approx \nrnd{5}{-23}(E^\nu/GeV)
 \int_{5E_{th}}^\infty \frac{d N_\nu(\txnd{cmb})}{dE^\nu_{cmb}} dE^\nu_{cmb} \,.  
\eeql

Naturally if the \zz \sp does not extend above threshold, which we will be here
taking as 5 MeV at \tx{cmb}, all such integrals should be set to zero.

\subsubsection{ Model with $\eb $ }\label{modeb}

 For general orientation we evaluate  \eq{tcb} in some of the
 schematic  models. In the model with a well defined \em \yyn,
we use \eq{nudistc} and integrate  from 5 MeV to infinity, one finds
 (with $x_{min}<<1$)
\beal{nudistf}
N_\gamma(\txnd{now}) &\approx& \frac{\nrnd{6}{15}}{sec^3}\frac{M_{pl}}{5 MeV}
 \nrnd{5}{-23}
(E^\nu/GeV) {\cal P}\\
\nonumber
&=& \frac{\nrnd{0.7}{15}}{sec^3}  (E^\nu/GeV) {\cal P}
~~~~~~~~~~~~~~~E^\nu>>5 MeV \, ,
\eeal

Thus a 10 m on a side  square  detector would have a rate 
$\sim 0.7 \times (E^\nu /GeV)  {\cal P}/ sec$. While much
depends of course on the values of ${\cal P}$ and $E^\nu $,
 this seems a  nontrivial result. We stress that this formula
results from taking  ${\cal P}$ constant, otherwise it
should be understood as some weighted  average
of  ${\cal P}$  over \em \tn s  $x$, according
to the earlier formulas.

\subsubsection{Spread Spectrum model}

From the discussion in subsection \ref{sprd}, there are two different
formulas to evaluate,  according to whether the
curve for  a given $E^\nu_{cmb}$ in Fig \ref{regions}
   is first stopped by the `escape' line or
by the $E_{max}$ line.
 
\subsubsection{$E_{max} \gtrsim  \tennd{15} GeV $  }

For high enough {$E_{max}$, above about \tend{15} GeV, still below
the planck scale,  all 
 \zz pulses have components which are stopped by the escape condition, 
  and so the second   option in
\eq{nudib} should be used for all \yys $E^\nu_{cmb}$ .

Dividing by $E$
for the number density we then  need the expression
\beal{2mod}
\int_{5E_{th}}^\infty \frac{d N_\nu(\txnd{cmb})}{dE^\nu_{cmb}}
dE^\nu_{cmb}\\
\nonumber
&=&\frac{\nrnd{6}{15}}{sec^3}  \frac{ M_{pl}}{E_{max}}{\cal P}
\int_{5 MeV}^{\infty} \bigl( (s\times E^\nu_{cmb}/GeV)^{-1/3}-1\bigr)
\frac{1}{E^\nu_{cmb}} dE^\nu_{cmb}\\
&\approx&
\nonumber
\frac{\nrnd{6}{15}}{sec^3} \frac{ M_{pl}}{E_{max}}{\cal P}
\times  3 (s\times 5\,MeV/GeV)^{-1/3}\\
\nonumber
&=&\frac{\nrnd{4}{21}}{sec^3}  \frac{ M_{pl}}{E_{max}}{\cal P}\,,
\eeal
and so for \eq{tcb}

\beal{tcc}
N_\gamma(\txnd{now})&\approx& \frac{\nrnd{5}{-23}}{sec^3}
(E^\nu/GeV) \times \nrnd{4}{21}\frac{ M_{pl}}{E_{max}}{\cal P} \\
\nonumber
&=&\frac{\nrnd{2}{-1}}{sec^3}(E^\nu/GeV)
\frac{ M_{pl}}{E_{max}}{\cal P}\\
\nonumber
&=&\frac{\nrnd{2}{18}}{sec^3}
\frac{E^\nu}{E_{max}}  {\cal P} \,.
\eeal

With ${E_{max}}$ so large and $E^\nu$ in the MeV or GeV range, this will be
very small in
comparison with \eq{nudistf}, reflecting
  the fact that only a small fraction of
the \zz pulse ``escapes".

\subsubsection{$E_{max} \lesssim  \tennd{15} GeV $  }

We now come to the case of the spread \sp where  ${E_{max}}$ is low enough
that most \zzs escape, using the first option of \eq{nudib}, so that one
has
\beql{nudisty}
\int_{5 E_{th}}^\infty \frac{d N_\nu(\txnd{cmb})}{dE^\nu_{cmb}}
dE^\nu_{cmb}
=\frac{\nrnd{6}{15}}{sec^3}
  \frac{M_{pl}}{5 \, MeV} {\cal P}
\, .
\eeql
which is  the same as in the first line
of  \eq{nudistf}, leading to the same conclusions.

\subsubsection{Emission Times}
Since a certain \zz \yy is required for the positron production, there are
limitations on how early the \zzs can originate. On
Fig.\ref{regions},  the descending curves indicate those \tn s for which a
certain \zz \yy appearing at \tx{cmb} can be emitted. Where such curves
enter the forbidden zone below the ascending straight line, 
 the \zzs can no longer ``escape." This earliest
time is given by the scaled time $x=(s\,E_{cmb}/GeV)^{2/3}$.
If we set $E_{cmb}= 5 MeV$ in order to have 1\, MeV at $z\approx 200$,
 this gives
an earliest \em  \t of about \nrd{5}{2} sec. If we set
  $E_{cmb}= 5\,GeV$ in order
to have a larger \csss for the positron  prodution,
 then the earliest \t becomes about \nrd{5}{4}  sec. Hence a
positron signal would indicate \bus activity around these \tn s or later.

\subsection{Reionizations by 511 keV Photons }
A flux of 511 keV \phos after recombination
 will lead to some reionizations of the newly formed hydrogen
atoms. Even though these \phos may not ``escape" to us, there is the
question  if these reionizations might be significant.

To examine this possibility we consider a certain density
$ N_\gamma(\txnd{cmb})$
  of 
the 511 keV \phos at \tx{cmb} and ask for the \prob for an hydrogen atom to
be subsequently ionized. 
 The \prob $Prob_{ioz}$
of an atom being ionized  per unit
\t   $dt$ is  
\beql{ion}
\frac{d\,Prob_{ioz}}{dt}=
N_\gamma(t) \, M\times \sigma_{ioz}\,,
\eeql
where $\sigma_{ioz}$ is the \csss for the ionization of an atom
by  a  511 keV \phon.   $M$ is a factor to account for the multiple
ionizations that can be induced by  a high \yy \phon.  

The density of the \phos will decrease with \t as 
$ N_\gamma(\txnd{cmb})\bigl(a(\txnd{cmb})/a(t)\bigr)^{3}
= N_\gamma(\txnd{cmb})\bigl(\txnd{cmb}/t\bigr)^{2}$ so that
 one has for the integrated \prob for an atom to be ionized 
\beql{tion}
 Prob_{ioz}\approx  N_\gamma(\txnd{cmb}) M  \sigma_{ioz}
 \int_{\txnd{cmb}}^\infty (\txnd{cmb}/t)^2 dt
= N_\gamma(\txnd{cmb})M\times \sigma_{ioz}\times  \txnd{cmb}\, .
\eeql
In principle $\sigma_{ioz}$ is also a function of \t due to 
the \rsh of the \phosn, but  the integral is dominated by \tn s
close to \tx{cmb}, so we may take it as constant.

Around   500 keV the \csss of \phos on hydrogen
is $\nrnd{2.8}{-25} cm^2$, see \eq{th}.
 Attributing this all to ionization, one has then for 
the  approximate  \prob for an atom to be ionized 
\beal{tiona}
 Prob_{ioz}&\approx& 
 N_\gamma(\txnd{cmb})M\times (2.8 \times 10^{-25} cm^2)\times
(9\times 10^{12} sec)\\
\nonumber
&=& \nrnd{7}{-2}  N_\gamma(\txnd{cmb}) M \, cm^3 \\
\nonumber
&=& \nrnd{3}{-33}  N_\gamma(\txnd{cmb}) M \, sec^3  \,.
\eeal

 Thus the density  $N_\gamma(\txnd{cmb})$ would have to  be very large
 to have a significant effect
 on recombination.
 For example, if we take the estimate \eq{nudistf} multiplied by \tend{9}
to refer to \tx{cmb}, 
with $1\,  GeV$ \zzs and $M=\nrnd{5}{4}$, as would correspond to converting
essentially  all
the \yy of an absorped \pho into ionization, one finds that the \prob of an
atom being ionized is $\sim \tennd{-4}\, {\cal P}$.

 Although this estimate will vary according to the form of $\cal P$, it
appears that the reonizations are not significant.  However, should it come to pass that
the positron mechanism turns out  to be much stronger
than in our simple estimates, then this  reionization
could lead to small ionized patches 
 after recombination,
which would be a  way of seeing the ``escaping'' neutrinos directly.
Also, in this case of very strong \zz pulses, it might be interesting
to consider $\nu+H\to \nu + e^- +p$ as a source of ionizations.
  Patchy reionization  could be a target for
 next generation ground-based 
  21cm and  CMB
 experiments which should be  capable of detecting it.
 These include improvements to HERA, which probes the 21cm power spectrum to $z\approx 8$ 
 \cite{Keller:2023ieg}, and 
 EDGES \cite{2017ApJ...847...64M},
 SARAS-3 
 \cite{Bevins:2022ajf}  and
 NenuFAR \cite{2024A&A...681A..62M}, 
 which set limits  on HI non-detection at $z\sim 20$.

\subsection{Conclusions on Positrons}

The above results suggest that when \yys of
 the \zzs in a pulse are low enough
so that most of them ``escape'', but high enough to produce positrons,
 then \eq{nudistf} will apply, which for ${\cal P}$
 not too small, can lead to a potentially 
 observable signal of $\sim (2\,or\,3)\, keV $
 x-rays at present.
In this
 energy range, the soft x-ray sky is dominated by local emission.
Observed fluxes are from diffuse gas
 in  the local hot bubble, the Milky Way
 diffuse hot gas {\cite{Ueda:2022gcv}} and
  even the circumgalactic hot gas \cite{Ponti:2022nix}.
  However it is possible that the characteristic form of the ``bump"
in the \sp   could allow it to be picked
out among these broad backgrounds.

\section{``Contained'' 
Bursts, the ``Heat Signal ''} 

 The \bus are  a non-equilibrium
phenomenon, quite different from the stages of quasi-thermal equilibrium
assumed in the  standard cosmology, and it is by this aspect 
that we hope to identify them.

Above we considered manifestations of \bu \pls that ``escape''.
We now consider  \bus that are ``contained'',
 where the \pls  interact  before \tx{cmb}. 
In particular  we examine 
 the possibility of nevertheless obtaining a signal from their
\yy deposit. If the volume occupied by the ``contained''\bu is relatively
small, it is possible that the deposited \yy leaves a significant signal. 
We  call this the ``heat signal''.

{ The  features of the heat signal, for our purposes, are
essentially  determined
by the  light cones of the general relativistic kinematics and not
 so much by any particular properties of the medium or
the \plsn. This is perhaps in contrast to the more familiar studies of
  diffuse  
 \yy injection   in the more local universe 
 \cite{Hu:1992dc}.  

  For the ``contained '' \bus, there are  two general mechanisms for the spread of the \bu \yy from
very early times. One is from the essentially free flight of the \zzs
before they interact. The
second is a diffusion of the \yyn, mostly carried by the electromagnetic and
strong interaction components, after the \zzs interact and  convert. Since diffusion is
a multiple \sctg process, one might  expect the first mechanism to usually
dominate. In this case, when the ``free flight'' determines
  the expected dimension, the
heated region will be simply ``frozen in'' at the free flight distance.
 However under certain conditions it may be that the diffusion distance
becomes significant. Below, we attempt
 a crude estimate as to when this may occur.

  There is a large body of later work, most notably the papers on detailed modelling of energy
transfer, pioneered by Chluba and colleagues,
 ranging from generic early energy injection
 \cite{Acharya:2021zhq, 2024MNRAS.527.9450A}.
 to hydrogen and helium recombination lines. \cite{Hart:2020voa}  and the synergy with   CMB anisotropies 
 \cite{Lucca:2019rxf}.
 The observational situation remains unsatisfactory with no progress on spectral distortion limits since the COBE-FIRAS experiment.
 The expectation is that only the implementation of  ESA's VOYAGE2050 roadmap will enable a definitive CMB spectral distortion mission, that will improve on FIRAS by some 5 orders of magnitude in sensitivity
 to the $\mu$ spectral distortion
\cite{ ESAVOYAGE2050}
  However even here the possible achievable angular resolution
 may be inadequate to probe the
 CMB parameter space of hot spots  that we seek to explore.

We stress that here  we are concerned with  direct heating of small
 regions soon after the thermalization of the CMB.

  We provide  some approximate,
 intuitive arguments for the relevant quantities. These
  could in principle be obtained
 from the global treatments such as the
 state-of-the-art Einstein-Boltzmann public
 codes such as  CLASS and CAMB that
 incorporate the full physics of diffusion damping.


 Given a \bu originating at 
 some spacetime coordinate $(t,{\bf x})$, the \yy
in freely flying \zzs will  spread
from this point along the light cone. 
After the \zzs have been stopped, the \yy will propagate much more slowly,
filling in the light cones. When  this further propagation
 is relatively  small
 the \yy is essentially contained within the ``free
flight''distance. As was explained above, this distance is governed  by 
\tx{free}. When the diffusive distance is greater the distance is governed 
by a dimension as in \eq{dfd}.  The contained \bus  
 are represented by the small circles in Fig.\ref{blc}.

Examples for complete ``containment'' of the \bus  may be seen
 in Table \ref{tab}
for those entries where $Prob_\infty \approx 0$. On Fig.\ref{regions} this
corresponds to \zzs emitted below the $x=sE$ line.  These figures are
for \zzs and are  based on an
estimate using weak interaction parameters.
 Other carriers of the propagating \yy
involving electromagnetic or strong interactions, will have a much more
localized containment of the \yyn.  However. when the \zzs carry nearly all
the \yyn, analogous to  the case of core-collapse supernovas, then the Table may
taken to apply realistically.  There is nevertheless the possibility that
some \pls of a \bu are ``contained'' while others are not, as in the 
spread \sp model.

 The signal  would then arise on 
 certain small angular scales on the CMB,
and this morphology would be  characteristic for the ``contained''  \busn.
The linear dimension at the CMB corresponding to ``free flight'' from
\tx{em} to \tx{cmb} was given in \eq{travb} as

\beql{travb1} 
 2\txnd{cmb}^{1/2}\bigl(\txnd{free}^{1/2}-\txnd{em}^{1/2}\bigr)
= 2\txnd{cmb}\bigl(x_{free}^{1/2}-x_{em}^{1/2}\bigr) \, , 
\eeql
using the scaled \t $x$ in the second writing.
 A signal would be on angular scales corresponding to this proper
size on the CMB.  The dimensional parameter in \eq{travb1} is $\txnd{cmb}$,
which expressed as a convential length is about $\tennd{4}$ Mpc. This is
the largest size a ``\bu spot'' could be (on the CMB)   and gets smaller
 as $x_{em}\to  x_{free}$, until the \bu ``escapes''.

 We recall that \eq{travb1} was found by using the
dimensions of the region in which a \zz pulse is absorbed, which is why
\tx{free} appears. This implies  the assumption that there
is no important  further spread of the \yy after the \zzs have been stopped,
which seems plausible in view of the small value of the diffusion distance
\eq{dfd}.

 However, this assumption could be wrong if a significant portion
of the \bu is transfered to a propagating form of \yy such as acoustic
waves.
 \eq{travb1} would approximately
   also apply to the propagation of sound waves, since
these  are  expected to travel at close to the speed of light in the
relativistic plasma, with \tx{free} replaced by a damping distance. The
dimensional parameter and so the size of the ``spot''should
 remain  essentially the same.

    This
damping was studied extensively in the work  
 on early
\yy injection mentioned above. For our considerations, we  note
that  according to \cite{Chluba:2012gq} for example, 
density perturbations in regions smaller than about $.02 Mpc= \nrnd{2}{4}pc$
 will be
damped out (by \pho diffusion)   by the time of the formation
 of the CMB. The wavelength of possible sound waves induced by  the \bus
might be expected, at most, to be the size of the \bu region at emission,
after \rsh to \tx{cmb}. This quantity is
 $ 9\times \tennd{9} (\txnd{em}/s)^{1/2}sec$. Since for the dimensional 
factor here  we have  $ 9\times \tennd{9} sec= 90 pc$, it appears
  that the part of the
\bu \yy appearing as density perturbations from sonic waves
will have been degraded  by \tx{cmb} and thus can also contribute to a ``heat
signal''.

While we imagine that the \yy of a \zz pulse will be dissipated mainly by a
diffusion-like process as in cosmic ray showers  \cite{segre}, the
subject of possible acoustic waves seems an interesting one in further
studies.

\subsection{Estimate of the Effect of  ``Contained'' Energy}\label{balhs}

 After the question of the size of a possible ``hot spot'',  the
remaining question is the magnitude of the  possible heating.

It is possible to make an estimate of the possible  significance of
this ``heat signal''  without  a   detailed description of the
thermalization process 
as developed in \cite{Chluba:2012gq}.

 Our argument is as follows.

 We gauge the  significance of this \yy deposit from a \bu
   by estimating what
\T rise $\Delta T$ it would produce, given the heat capacity $C$ of the
region of ``containment''. If this $\Delta T$ turns out to 
be not very small compared to the \T at \tx{cmb}, it  suggests a
potentially observable effect. This is a conservative estimate, since the use
of the  thermal heat capacity assumes the deposited \yy has been spread over
all degrees of freedom uniformly, while it is possible that the \yy
does not have \t to completely thermalize.

\subsubsection{``Free Flight''Region}

We first consider the heating of a region whose volume is determined by  the
length \eq{travb1}.
 One   needs  the \bu \yyn,
after the \rsh to \tx{cmb}. From our assumption that the \yy of a \bu at \em
is the \yy in a horizon volume  or $\yy\, density \times volume  $, we write

\beql{efree}
E^{\bu}_{\txnd{cmb}}=
U_{em} V_{em}\frac{a(\txnd{em})}{a(\txnd{cmb})},
\eeql 
with $U_{em}$ the \yy density at \em and $ V_{em}$ the horizon volume
at \emn.
We find the $\Delta T$ induced by this \yy from the heat capacity at 
\tx{cmb}. In terms of $C_V$, the heat capacity per unit volume, we have
at \tx{cmb} 
\beql{c1}
C= C_V\times V_{cmb}= \frac{4 U_{cmb}}{T_{cmb}}\times V_{cmb}\, ,
\eeql
where in the second step
 we use the relation between $C_V$ and $U$ for a relativistic gas,
$C_V=4\,U/T$.
  $V_{cmb}$ is the volume occupied by the \bu at $\txnd{cmb}$.
Taking the ratio of \eq{efree} to \eq{c1} and dividing by $T_{cmb}$
we find at \tx{cmb}
\beql{rat}
\frac{\Delta T}{T}=\frac{1}{4}~  \frac {U_{em}}
{U_{cmb}}
     \frac {a(\txnd{em})}{a(\txnd{cmb})} \frac{V_{em}}{V_{cmb}}.
\eeql
The ratio of the \yy densities is as $(T_{em}/ T_{cmb})^4$ or
 $(a(t_{cmb})/ a(t_{em}))^{-4}$  so \eq{rat} is
\beql{rat1}
\frac{\Delta T}{T}=\frac{1}{4} 
     \biggl(\frac {a(\txnd{cmb})}{a(\txnd{em})}\biggr)^3 \frac{V_{em}}{V_{cmb}}.
\eeql

The ratio of  volumes goes as the cube of the linear dimensions. At \tx{em}
the linear dimension is the horizon or  $2\txnd{em}$, and for the \bu
at \tx{cmb} we have from \eq{travb} the linear dimension approximately
$2\txnd{cmb}^{1/2}\txnd{free}^{1/2}$
Inserting the cubes in \eq{rat1}

\beql{rat2}
\frac{\Delta T}{T}=\frac{1}{4} 
     \biggl(\frac {a(\txnd{cmb})}{a(\txnd{em})}\biggr)^3 
\biggl(\frac{\txnd{em}}{(\txnd{cmb}\txnd{free})^{1/2}}\biggr)^3
= \frac{1}{4}~\biggl(\frac{\txnd{em}}{\txnd{free}}\biggr)^{3/2}
 = \frac{1}{4}~ \kappa^{-3/2}\,,
\eeql
using $a\sim t^{1/2}$.
(If one does not wish to make the approximation
$\txnd{em}/\txnd{free}<<1$
in the linear dimension  then there is an
additional factor $\bigl(1- \txnd{em}^{1/2}/\txnd{free}^{1/2}\bigr )^{-3}$.)

}

We are considering the `'contained'' case so  $\kappa=\txnd{free}/\txnd{em}$ 
is greater than one. But 
for \bus that are ``well contained'', where $\kappa$ is not very much
larger than one, it thus  appears that an interesting degree
 of ``heating'' is possible. 
 This signal  would be visible  as  a deviation from a planckian \sp
in regions of size \eq{travb} or \eq{dfd}
if the \yy has not had \t to thermalize.  Or if the \yy is  thermalized,
 as a simple  \T fluctuation, In this latter case, it would be necessary
to disentangle this from other sources of \T fluctuations.
Understanding whether essentially complete thermalization  occurs 
or not would require a
 detailed analysis going beyond our
simple one-component model of the plasma.
But  the nonplanckian \spn, which of course
is a nonequilibrium feature, would be characteristic
for the \busn .

\subsubsection{ ``Diffusive'' Region }\label{diffspread}

{ The above estimate for the heat signal  was based on
using the ``free flight"  estimate for the size of the heated 
region at \tx{cmb}. We can do the same for the opposite case,
 for the small ``diffusive ''
region, using the estimate \eq{dfd}  for the size of the region.

 Since we  argued  that  
  the dimensions of the `diffusive
  region are constant, independent of \em \tn, the
only variable in the problem is the \yy of the \bun. 
As expressed in \eq{buea} in terms  of  the scaled \em  \t $x$ this is:
$ E^{tot}_{cmb}= \nrnd{2}{56}\, x^{3/2}\,M_{pl}$

One then finds
\beql{difsig}
\frac{\Delta T}{T}\sim \frac {\nrnd{2}{56}\, x^{3/2}\,M_{pl}}{T^4 d^3 ly^3}
\approx
  \nrnd{7}{18} \frac{x^{3/2}}{d^3}=0.3\, (\txnd{em}/sec)^{3/2}\frac{1}{d^3}\,.
\eeql

The quantity $d$ is the linear
dimension of the potentially heated region in  light years. $T$ is
the \T at \tx{cmb}.

 If other propagating modes such as sound waves are important  
in the \yy spread, then one may expect an effect in-between the two extremes
we have examined here: for a large ``ballistc region''
and a small ``diffusive region''.

Finally, we note we have  been  assuming
 that the full \yy of the \bu is
 ``contained''. However, 
for \bus that are not very early, or which contain a wide spread of \pl
\yysn, this may not be the case. For example, in the simple spread-out model
of section \ref{sprd}. \eq{mxesa}, the fraction of the \yy ``escaping''  is
 $(x_{em}/s)(GeV/E_{max})=
 1 \times (t_{em}/sec) (GeV/E_{max})$.
In such cases the  $\Delta T/T$ value must be reduced according to the fraction
not escaping. However for high \pl  \yys or early \em times the escape fraction  goes to
zero and all the \yy is ``contained''.  }

\section{Many Small Bursts}

In the previous section, we
 studied the heat signal from a distinct single
\bun. As we go back in \tn, to ever smaller $x$, 
although the \bu get weaker, $\sim
x^{3/2}$, the
great  increase in the number of potential emitting regions may compensate
 for the decreasing intensity.
If the \bus do in fact arise from transitions
 to new spacetime regions, it is
plausible that the underlying quantum tunnelling proceeds with higher \prob
for the smaller horizon-sized regions at early \tn s as $x\to 0$.

In this section, we thus speculate on the possibility of many small
overlapping \bu regions. We assume that all \yys from a \bu are
``contained'' and so the \yy delivered to \tx{cmb} from a single \bu is
\eq{buea}, namely $\nrnd{2}{56}\, x_{em}^{3/2}\,M_{pl}.$

\subsection{Total Energy from Very Early Bursts}

According to \eq{econtrb}, the \yy density contribution from such many early
\t \bus would be the integral of $\nrnd{3}{15}\frac{M_{pl}}{sec^3}\frac{{\cal
P}(x)}{x}dx$. If ${\cal P}(0)$  does not vanish we expect a logarithmic
divergence for the \bu \yy density
\beql{logmin}
\epsilon_{\bus} \sim
\nrnd{3}{15}\frac{M_{pl}}{sec^3}<\Delta \, x{\cal P}(0)> (-ln\, x_{min}),
\eeql 
where $< \Delta x  {\cal P}(0)>$ is some typical value of the integral
 $\int {\cal P} dx $  and $x_{min}$ could correspond
to the planck \tn, $(x_{min})\sim \tennd{-56}$.

Since we have many independent \bus from all directions, these would give a
uniform background on the CMB. Taking the log to give one or two orders of
magnitude, this suggest an \yy density of \bus 
\beql{lnden}
\epsilon_{\bus} \sim
(\tennd{16}-\tennd{17})<\Delta x {\cal P}(0)> \frac{M_{pl}}{\rm sec^3}
\eeql
As remarked following \eq{budensbu} with ${\cal P}(0)$ not too small
this could be enough to give the 
   total radiant energy density   of the
universe. This is not surprising since we have
 assumed that the \bus  are generated at  the horizon scale. 
However we stress  that a basic assumption in this paper is 
that $\cal P$ is so  small
that the usual cosmology is unaffected. It might be interesting to drop
this assumption.  This would invite a  much wider discussion.

\subsection{Fluctuations}

  \eq{logmin} represents simply a uniform average
 density, { and so of itself is not obviously observable.
However, what could be more characteristic
for the picture  are  the associated statistical
fluctuations. 
With many independent very early
 \bus of small \yy one
expects a statistical distribution of the \bu \yy over the patches of the sky
under observation.

We consider the quantity   $\overline{\delta E^2}$ or the variance
characterizing the \yy fluctuations from the \busn, whose square root
is the typical \yy fluctuation.
Since we take the \bus to be independent statistical events a variance
will result from the fluctuations in the number of events
in the sample, as for a poissonian or gaussian distribution, where
$variance \sim {n}$, where $n$ is the average number of events. That is, we
interpret the $n$ in \eq{nobs} to refer to the  average number of events,
which then gives the variance.
There could be  further sources of variance
such as   fluctuations in the \bu \yy from
a given epoch, but here we will simply take $\overline{\delta E^2}
=\overline{\delta n^2} E^2=n E^2$.
 For an interval of scaled cosmic time $dx$ one  has 
$n= \frac{dn}{dx} dx$

That is,
the contribution from an interval of scaled cosmic time $dx$ will be 
\beql{patch}
\overline{(\delta E)^2}=\frac{dn}{dx}E^2(x)\,dx\,, 
\eeql
where $E(t)$ is  the \yy in a single \bun,  \eq{buea},
namely $\nrnd{2}{56}\, x^{3/2}\,M_{pl}$.
We take $\frac{dn}{dx}$ from \eq{nobs} so that \eq{patch} becomes 
\beal{nobsa}
\overline{(\delta E)^2}&=&
 V_{obs}\times\frac{\nrnd{2}{-41}}{sec^3} 
 \frac{{\cal P}(x)}{x^{5/2}} dx \times \biggl(\nrnd{2}{56}\,
 x^{3/2}\,M_{pl} \biggr)^2 \\
\nonumber
&=& \frac{V_{obs}}{sec^3}{\nrnd{5}{71}{\cal P}(x)}\,x^{1/2} M^2_{pl} dx.
\eeal
{ Adding up the contributions from all the $dx$ intervals, one has for the}
 integrated \yy fluctuation itself then
\beal{nobsb}
\bigl(\overline{(\delta E)^2}\bigr)^{1/2}=
\bigl(\frac{ V_{obs}}{sec^3}\bigr)^{1/2} \nrnd{7}{35}\,M_{pl}
\biggl(\int dx\, {\cal P}(x)\,x^{1/2}  \biggr)^{1/2}.
\eeal
An interesting aspect of this formula is that if ${\cal P}$
is a  not too strongly increasing
 function for $x\to 0$, then the decreasing $x^{1/2}$
could lead to  a peak at some $x$,  designating an epoch giving the
 greatest  fluctuations.

Finally we have for the relative fluctuation of the \bu \yy
at \tx{cmb}, using \eq{econtrb} for the total \bu \yy density
$\int \nrnd{3.2}{15}\frac{M_{pl}}{sec^3}\frac{{\cal P}(x)}{x}dx$
\beal{nobsc}
\frac{\bigl(\overline{(\delta E)^2}\bigr)^{1/2}}{E}=
\bigl(\frac{ V_{obs}}{sec^3}\bigr)^{-1/2} \nrnd{2}{20}
\frac{\biggl(\int{\cal P}(x)\,x^{1/2}dx  \biggr)^{1/2}}
{\int\frac{{\cal P}(x)}{x}dx}.
\eeal
With the sort of values for $ V_{obs}$ mentioned after \eq{nobs}, the
prefactor here is relatively large, and could be made larger by
observations with higher
angular resolution.  There is also the large value
of $1/\surd{\cal P}$ implicit in the ratio factor with small $ {\cal P}$.
 The observational aspect of  many independent small fluctuations
could thus be that the non-equilibrium ``heat signal'' features,
 discussed above for single \busn, become more pronounced for samples with
higher angular \rsln.

 However since  $x$ is always less than one,  these factors favoring large
fluctuations will tend to  be compensated 
by  the small value for the ratio of the integrals. For example, with 
${\cal P}(o)$ nonzero there is the behaviour $x^{3/4}/lnx $ for $x\to 0$,
which would give 
a factor $\sim \tennd{-44} $ towards the planck time.

\section{Time-dependent Phenomena} \label{tdep}

\subsection{Intrinsic  Burst  Length}
One of the most intriguing possibilities concerning the \bus would be the
observation of an arriving   \bun.    Much as we see the brightening 
  of a star to a supernova,  a small spot on the CMB might brighten
and then fade. But  unfortunately such an observation seems unlikely, as we
discuss in this section.

  In  Eq 4 of \cite{jl}, we took the intrinsic
 length of a \bu  to be given by the size
of the emitting region, 
 using the horizon size.  With the \rsh to the present
   the length of a \bu  if it could be directly seen at the
 earth  would be
\beql{budur} 
9\times \tennd{9} (\txnd{em}/s)^{1/2} sec
=\nrnd{3}{15}x^{1/2}_{em}sec\,.
\eeql
  We note that this estimate of the time scale is based on using
the horizon size as giving the natural dimensions. If there is a period
of ``\zz trapping'' \cite{schramm}  as in core-collapse supernovas,
the \t scale could be somewhat longer.

For early \em \tn s, $\txnd{em}<<1 sec$ 
it thus seems conceivable that with
an observation period of years the ``arrival'' of a \bu could be seen, in
principle.
  However, even if there were \pls that would `'escape'' from such
early \tn s, their very great \rsh would make them  undetectable,  at
least as pulses.
We thus   consider our
  indirect methods of \dtn on the CMB, where however 
 a number of points concerning \t dependence should be considered.

\subsection{Movement of our Backward Light Cone}

A preliminary question regarding  \t dependence concerns
  the movement of our BLC, the
increase in the radius of the large circle in Fig.\ref{blc}. After a
sufficiently long time, an observation in any  given direction
will be looking at a different patch at \tx{cmb}, with a different \bu
history. However,
 the time spread or averaging  for  a given
observational arrangement  is $d_{thom}$,
 or several thousand  years (\eq{h2}).
 On the other hand at \tx{cmb}, the BLC  moves out
 $a(\txnd{cmb}) \times 1\,\, ly $ in a year,
or \tend{-3} ly in one earth year. Therefore,
on the scale of   years for an  observation, one is always effectively  viewing
the same  region of the CMB, and  there should be little change  in the  
 phenomena originating from the \busn.
(But it might be worth good record keeping of present
observations for the use of colleagues in a distant future.)

\subsection{Delay Due to Signal Production}

An issue is the \t delay connected with the production of the
observable signal. The production of the signal can  involve some  \tn,
 and we consider this for
the two mechanisms, the ``positron signal" and the ``heat signal''.

\subsubsection{Positron Mechanism}

Here an essential element in the \t structure will be the \t it takes for
a positron to annihilate after it is produced. At the low density 
of electrons around \tx{cmb} or later,  the positrons will wander for a considerable 
\t before annihilating. We may roughly expect this to be on the same order
as that determined by $d_{thom}$ or several thousand ly. The lifetime
for positrons in ordinary  materials is $\sim \tennd{-10}$sec. Rescaling for the
density of electrons at \tx{cmb} one finds $\sim \tennd{11}$sec.
 Thus for observations made 
on the scale of years, any initial \t structure for the 
positron signal will be washed out,  even not accounting for the \rsh
to earth \tn.

\subsubsection{Heat Signal}\label{hsd}

We consider the simple cases of a fully contained \bun, symbolized
by the smallest circles in Fig.\ref{blc}. The small circles are 
to be imagined in or on the finite thickness of the large circle.
 The heat signal 
arises in the region in which the \bu was  absorped,  given
essentially by \tx{free}. According to section \ref{cntnd} the
{ largest} size of this
region  at \tx{cmb} is given approximately by 
\beal{travc} 
size_{cmb}\approx 2\txnd{cmb}^{1/2}\txnd{free}^{1/2},
\eeal
which can vary greatly, according to the value of \tx{free}.

Generally, there are  two  possible sources of a  \t dependence of the signal.
  One is from  changes in the nature of the ``spot'' itself during an observation
period.
The second could be alterations in the ``observability'', 
 due to  the motion of the
 large circle in Fig.\ref{blc} during the observation time. Over a long
enough period of \tn, a ``spot'' which is  observable could  move on or
 off  the big circle.
However, both types of effects should be small, not least because of the \rsh
of the observation \t at earth to \tx{cmb}.
 We characterize the length of an observation period in terms of
$\Delta t_{obs}$ earth years, which corresponds to $\nrnd{1}{-3}\Delta t_{obs}$
 years at \tx{cmb}.

As an example of the first type of \t dependence we  consider the size of
the \bu region. Due to the expansion,  the ``spot'' is growing during
the observation. Other effects of this type might be shifts in the \sp
under study, but this must await a detailed study of the thermalization
process.

To estimate the change in size, we use \eq{travc} taken as a function of
$\txnd{cmb}$.   Up till now, 
we have treated $\txnd{cmb}$ as a constant, but this factor
in \eq{travc} arose from the expansion factor $a(\txnd{cmb})$, and the
expansion continues during the observation period. We then find for the
relative size change 
\beql{szch}
\frac{\Delta size_{cmb}}{size_{cmb}}=\hf \frac{\Delta \txnd{cmb}}{\txnd{cmb}}=
\hf \nrnd{1}{-3}  \frac{\Delta t_{obs}}{\txnd{cmb}}.
\eeql
Since $\txnd{cmb}\sim \nrnd{4}{5} \rm yr$ one sees that an observation 
 this type, with years duration, will have
to contend with the small factor $\sim \tennd{-8}$.

Coming to the second type of effect, due to the movement of our
BLC, the small parameter is the  movement.
 In a $\Delta t_{obs}$ of   one earth year, the radius
of the large circle in Fig. \ref{blc}
 increases by $a(\txnd{cmb})\Delta t_{obs}\approx \tennd{-3}\rm yr. $
For example, if a small \bu is near
the outer edge of the observation band,
 in one year it will move $\sim \tennd{-3}\rm yr $ 
closer to the center, brightening slightly.
This is to be compared with the width of the observation band, symbolized
by the thickness of the big circle. This dimension is given
by $d_{thom}$ or several thousand years. The ratio then is also
 small $\sim\tennd{-7}$.
 It would seem that such effects
would be strongest when the size \eq{travc} is about the same of the width
of the band. Otherwise for small sizes the \t variation will be controlled
by the variation of the observability of the signal under consideration
across the band, involving a detailed analysis.

 A  important practical
consideration here is likely to be if the sought-for deviations from the
thermal \sp are  detectable  within  the relatively short observation
 \tn s.
 From \eq{budur}, it seems
that \bus that somehow could be detected directly at the earth might have
a reasonable  length in \tn.. However, we conclude
here  that those observed via the CMB may
be regarded as permanently fixed on the sky, up to the very difficult
measurements just described.

\section{Assumptions,  Approximations, Further 
Work} \label{ass}

In this section we note some of our most important
 assumptions and approximations, with the implied suggestions for
further work.

 Our most salient  assumption is that the
 presently standard cosmology, with its parameters,  may
be used in the presence of the \busn. This seems a quite plausible assumption
when ${\cal P}$  the \bu  \prob is very small. But  how large a  ${\cal P}$
is permitted within this assumption is a complex question. One simple
estimate for its validity
would follow from requiring that the \bu \yy transmitted to
some relatively  recent times, say
\tx{cmb}, be much less the known \yy at that \tn. According to
\eq{econtrb} the \bu \yy density at \tx{cmb} is the integral of
$\nrnd{3}{15}\frac{M_{pl}}{sec^3}\frac{{\cal P}(x)}{x}dx$. With 
  the  standard \yy density
 $(\pi^2/15)T^4=\nrnd{8}{33}GeV/sec^3$ at \tx{cmb},
 this leads  to the amusingly simple condition

\beql{plim}
\int_0^1 {\cal P}(x)\frac{dx}{x} <<1\, . 
\eeql

 While it seems evident that \eq{plim} is a necessary condition that
 the standard cosmolgy is not significantly affected, an interesting 
subject for further study is if it is also a sufficient condition. 

 A further assumption was that in writing the simple \eq{escale} for the \rsh of the total
 \yy of a \bu we have assumed that the \rsh is radiation-like, even for the
secondary reaction products in the medium. This assumption might have to
be reexamined in situations where a significant portion of the \yy ends up
being carried by nonrelativistic components.  

On a more technical level, one of our simplifying assumptions is the use of the
 radiation dominated epoch  formula $a\sim t^{1/2}$
all the way to \tx{cmb}, while strictly speaking it should only be used
until the somewhat earlier \t of radiation-matter equality \tx{eq}.
This however does not lead to a very large error in the numerical
estimates; the value of $a(\txnd{cmb})$ this way is \nrd{0.7}{-3}
 instead  of the more correct
\nrd{1.0}{-3}, and  it permits much more transparent formulas.
 and  identification of  the essential issues.

\subsection{Neutrino Absorption Parameters}
The use of only one degree of freedom for the 
ambient medium and a generic \zz \csss to find the \zz absorption, allows
us a simple identification of the main features of our problem.
 A more exact treatment would take into account the increase in the number
of degrees of freedom with \yy and particular features affecting the 
\csss such as nucleosynthesis or phase transitions.
While the general features we have found should remain, alterations in the
numerical factors may be expected.

\subsection{``New Physics"}\label{new}

A further  assumption was our use of arguments and parameters
 based on a simple  extrapolation of
presently known information, while there is the  possibility
of ``new physics'' at yet higher \yys than have been explored
todate. 

The most likely of these possibilities  concerns our \eq{cs}, which controls our estimates
of \zz absorption. This formula represents lowest order perturbation theory
for weak interactions, which gives the \csss increasing with
center-of-mass  \yy squared $S$. It is not
plausible that this increase continues indefinitely at very high \yyn. Once the
\csss reaches a value corresponding to the geometric range of the
interaction, unitarity arguments suggest that the growth must saturate.  The
linear growth with $S$ will probably then go over to the much
slower  $(ln S)^2$ growth,
as  is known for strong interactions \cite{leo}. This geometric range of the
 weak interactions is given by the exchange of $W$ and $Z$ vector bosons,
that is, by our $M= 100\, \rm GeV$ parameter via $range=1/M$. We thus
 anticipate that our estimates are
approximately valid until \yys where
\beql{unt}
\biggl(\frac{\alpha}{M^2}\biggr)^2 S\approx \pi \bigl(\frac{1}{M}\bigr)^2\,  
\eeql  
or until the center-of-mass \yy squared
\beql{cme}
S=\frac{\pi}{\alpha^2} M^2\approx \nrnd{6}{8} GeV^2 \,.
\eeql
 In
terms of the interaction of our \bu \pls of \yy $E_{em}$
 with the ambient plasma at \T $T$
this means that  our considerations are valid up to  values of
 $E_{em}\times T$
given by $E_{em}T=  E_{em} \nrnd{2.5}{-4}eV/a)\approx  \nrnd{6}{8}GeV^2$.
(For the kinematics see the discussion following \eq{cs}).

 Using $a=(t/\txnd{rad})^{1/2}$
this implies that our simple discussions apply when
\beql{apply}
E_{em}<(t_{em}/sec)^{1/2}\nrnd{5}{11} \rm GeV \, .
\eeql

This constraint will only be relevant for the highest \yysn;
the values  in the Table are unaffected, for example.
But it could  play a role
 towards the planck scale 
 $\sim\tennd{19}\,GeV,$ if we wish to speculate about that regime.

 One may rearange \eq{apply} in terms of the scaled \t $x$
to give the condition
\beql{applya}
x> \tennd{-36} (E_{em}/GeV)^2 \, .
\eeql
On Fig.\ref{regions} the boundaryb of this region would be  a quadratically 
increasing curve with  very small coefficient. It finally  crosses
the linearly increasing ``escape'' boundary, but only at very high \yyn,
near the planck scale. This implies that our treatment remains qualitatively
correct for essentially the entire region of the $(E_{em}, x)$  plane, except
for planck scale \yys at late \tn s.

\subsection{Calculation of $\cal P$}

The most important open question in our presentation lies in
the production of the \busn.  With knowledge or models for
 $\cal P$, its size  
  and  \t dependence, and of how the potentially
 visible \yy of the \bu is set free, our
formulas could be evaluated in more detail. Particularly
useful would be a knowledge of the \pl \yy \spn , to avoid the very simplified
model  assumptions of section \ref{ill}.

It should also be noted that in addition to \bus originating
from \gvtl processes like the creation of
 ``baby universes'' or supermassive \bh formation,
it has been suggested that there could have been `'little bangs'', connected 
with phase transitions in the very early universe \cite{kort}. It is to be
anticipated that these also will lead to the kind of signals we discuss.
Further work here  would  be
 the adaptation of our  parameters  for these kinds of processes.

\subsection{Gravitational Waves}
An important associated question would  be the production
of { gravitational}  waves, which might 
 then be an observable signal for the \busn.
The recent discovery of a 
stochastic { gravitational} wave background \cite{nanograv}, 
first interpreted as a signature of 
{ 
the gradual inspirals preceding early supermassive black hole mergers, }
 although such events rates may fall short
 by up to an order of magnitude \cite{zaldarriaga2024}, 
 might be equally attributable to other  physics \cite{newphysics}
  that could include phenomena  such as early bursts. 
It would be of great interest if the angular resolution of 
 gravitational wave 
\dtn allowed a directional  identification with one of  our ``heat signals''.

\subsection{Burst Thermalization}
A  detailed description of the partial thermalization of the \bu
\yyn, particularly that of the \zz pulse, would be valuable in connection with
the appearance of the nonequilibrium ``heat signal''. Knowledge of the 
form of the deviation from the standard planckian \sp would be important.
 Due to the increase
of the \zz \csss with \yy it is likely that this deviation
 appears as a high \yy
component, but a detailed calculation would be useful.
{ In addition to
determining more accurately the diffusion length, such an analysis should
also examine the free flight length,  for a more detailed understanding
of the heat signal.

An aspect of further work on this topic  could be the
study of collective mechanisms such as plasma oscillations and acoustic    
waves for the spread and dissipation of the \bu \yy \cite{Chluba:2012gq}. These will of course
involve much stronger interactions than for \zzs and so smaller regions.  But
if they are important and do not dissipate,
these would cause hot spots in the CMB and generate non-gaussian
features on small angular scales in the centers of the \busn. See for example statistical discussions
of using such features to probe CMB non-gaussianity
\cite{Chingangbam et al.(2012)} and more recently by \cite{Khan and
Saha(2022)}.

Finally, an important aspect
 of the ``heat signal'' which would be enabled by detailed modeling
 is the contribution of the \bus to the 
 power \sp of the CMB temperature fluctuations, in effect via  a generalization of
 peak statistics in the cosmological context \cite{1986ApJ...304...15B}.
 Roughly speaking, the size of the contribution will be given by the
magnitude of $\cal P$ and the angular scale will be governed by \tx{free}, or
$d$.  By  comparison with existing or
future data, limits on or values for these quantities could be estimated.}

\section{Conclusions}

It is possible, and we feel likely, that major transformations
 of spacetime in the very early universe lead to peripheral phenomena,
which we call ``\bus''\cite{jl},  that 
 are potentially  observable in our part of the universe. 
 
We have identified three possible signals of such \busn. One, and probably
the most difficult observationally,
 would be a nonthermal addition to the low \yy \zz \sp anticipated for
thermal relic  \zzs (subsection \ref{eab}).
The  two others  would be
effects conceivably accesible by present technologies or their
 extensions , namely small regions on the CMB with a nonthermal \pho \sp
and finally, the production of positrons
 leading to  a $\sim (2\,or\,3) keV$ soft
 x-ray population (section \ref{epl}).
While none of these signals seem easy to detect, their \dtn would
 would  provide evidence for the most dramatic events in the history
of the universe   and 
 usher  in a new era of observational cosmology.

\section{Appendix,   Units}
 In this paper we continue to  use the simplest FRW cosmology
  as explained in the appendix of \cite{jl}, with the notation and
parameters described there. However, the parameter $t_{now}$, which appears
in the late-time $a(t)=(t/t_{now})^{2/3}$
 should, in
view of more recent data, be set to $\txnd{now}\approx \nrnd{2.9}{17}$ seconds. 
To the one-place accuracy we are using, the other  parameters remain
unchanged, so we 
 thus use $a=(t/\txnd{rad})^{1/2}$ for early \tn s with
$t_{rad}=1.9\times10^{19} sec $. Also we use
 $a(\txnd{cmb})=1.0\times 10^{-3} $

Additional  parameters we use  are $t_{cmb}=\nrnd{9}{12} sec$
 so that
$\txnd{cmb}/\txnd{rad}=\nrnd{4.7}{-7}$,
 and for the radius
of our Backward Light Cone at \tx{cmb}, $R_{cmb}(blc)\approx \nrnd{7}{14}
sec$.
 Also  $T_{now}=2.9K=\nrnd{2.4}{-4} eV$ and  $T_{now}^{-1}=\nrnd{2.8}{-12}
sec$.

 We generally use natural units ($\hbar =1, c=1$, k=1)  and so
 express distances in light-seconds or light-years and \T in eV.


\begin{thebibliography}{00}
\bibitem{nist}
See  the ``x-ray mass attenuation tables" of NIST,
available at 
 'physics.nist.gov/cgi-bin/Xcom/xcom3\_1'
We consider only hydrogen, the contribution from
 helium may be seen to be small.
\end{thebibliography}
\end{document}